# *Mechanical metamaterials: does toughness characterize fracture?*

**Authors:** Angkur Shaikeea [1*], Huachen Cui [2,3*], Mark O'Masta [1,4], Xiaoyu (Rayne) Zheng [2,3§] and Vikram Deshpande [1§]

**Author Affiliations:**

[1]Department of Engineering, University of Cambridge, Cambridge, CB21PZ, UK

[2]Advanced Manufacturing and Metamaterials Laboratory, Department of Mechanical Engineering, Virginia Tech, Blacksburg, VA 24060

[3]Advanced Manufacturing and Metamaterials Laboratory, Civil and Environmental Engineering, University of California, Los Angeles, CA 90095

[4]HRL Laboratories, LLC, 3011 Malibu Canyon Road, Malibu, CA 90265, USA

[*] These authors contributed equally to this work

[§]**Corresponding authors:** V.S. Deshpande    Email: vsd@eng.cam.ac.uk

                                X. Zheng       Email: rayne@seas.ucla.edu

*Summary paragraph*

*Rapid progress in additive manufacturing methods has created a new class of ultralight and strong architected metamaterials that resemble periodic truss structures. The mechanical performance of these metamaterials with a very large number of unit cells is ultimately limited by their tolerance to damage and defects, but an understanding of this sensitivity has remained elusive. Using a stretching-dominated micro-architecture and metamaterial specimens comprising millions of unit-cells we show that not only is the stress intensity factor, as used in conventional elastic fracture mechanics, insufficient to characterize fracture but also that conventional fracture testing protocols are inadequate. Via a combination of numerical calculations and asymptotic analyses, we extend the ideas of fracture mechanics and develop a general test and design protocol for the failure of metamaterials.*

**Introduction**

Lightweight, three-dimensional micro- and nano- lattices represent a promising class of low-density materials ($< 100 \text{ kg m}^{-3}$), with numerous applications like thermal insulation *(1)*, battery electrodes *(2)* and energy absorption *(3)* to name a few. They achieve remarkable mechanical and functional properties *(4-6)*, governed in part by the parent material and in part by their architecture and hence commonly referred to as mechanical metamaterials. Recently, additive manufacturing techniques like projection microstereolithography *(7)* and two-photon lithography *(8)*, have led to realization of polymeric *(4)*, metallic *(5)* and ceramic *(9)* metamaterials. Effective properties have also been defined for these materials, with extensive reporting of their modulus and compressive behavior *(5, 10-13)*. As manufacturing processes scale-up resulting in the proliferation of metamaterials, an assessment of their tolerance to manufacturing defects and damage is critical. In continuum solids, the material property known as fracture toughness is used for such an assessment. However, an understanding and measurement of toughness in three-dimensional (3D) metamaterials has remained elusive partly due to the inability to-date to manufacture samples with sufficient number of periodic unit cells *(14-17)* so as to generate the required crack-tip $K$-field.

In a continuum elastic material, a stress distribution known as a $K$-field is established near the crack tip *(18)*, which is described by a single scalar parameter known as the stress intensity factor $K_I$. Rapid, unstable crack advance will occur when $K_I = K_{IC}$, where $K_{IC}$ is the fracture toughness of the material *(18)*. All previous work *(14-17,19-22)* has assumed a-priori that these ideas extend to truss type 3D metamaterials and used well-established experimental *(23)* and theoretical *(18)* methods to estimate $K_{IC}$. Here we show via experimental measurements and numerical analyses that neither can fracture of such metamaterials be described solely by $K_{IC}$ but also that standard measurement protocols, developed for continua such as metals and ceramics, are inappropriate for architected materials.

Our study used a stretch-dominated *(12)* metamaterial, comprising a network of struts where the nodes are arranged in a face centred cubic (fcc) lattice - the so-called octet-truss *(13,24)*. Fabrication of large 3D specimens, comprising of nearly 10 million periodic cells and cell sizes as small as 150 µm, was accomplished using a novel large area projection micro-stereolithography (SLA) system where each layer is fabricated via a continuously moving spotlight so curing occurs in sub-sections. This enables the creation of metamaterial specimens with millions of unit cells and homogeneous mechanical properties *(7)*. (*Methods* and Fig. S1 as well as SI including Figs. S2-S3 for details of imperfections and material properties). The system utilizes a dynamic, co-ordinated optical scanning system to continuously project and scan two-dimensional (2D) slices onto an ultra-violet (UV) curable photopolymer to produce large-scale three-dimensional (3D) parts with micro-scale resolutions. This enables us to manufacture specimens with embedded cracks and with a sufficient number of unit cells so as to develop a $K$-field and thereby perform a range of valid fracture toughness measurements. The results testify that the finite unit cell sizes restrict the development of singular stresses, with the consequence that $K_{IC}$ alone is an insufficient parameter to characterize fracture. Via a combination of measurements, including in-situ X-ray computed tomographic (XCT) observation and large-scale numerical simulations (with billions of degrees of freedom), we illustrate that a combination of $K_I$ and the so-called T-stress fully characterizes the fracture mechanisms over a wide range of specimen densities, parent materials, cell sizes, crack sizes as well as loading configurations.

**Results**

*Is stress intensity factor a sufficient parameter to set the failure mode?*

Cubic octet-truss specimens of dimension $2W = 2H = 2B = 30$ mm (Fig. 1a) were fabricated with trimethylolpropane triacrylate (TMPTA) as the parent material. This polymer is linear elastic with a modulus $E_s \approx 430$ MPa and fails in a brittle manner with failure strength $\sigma_f \approx 11$ MPa (failure strain $\varepsilon_f \approx 0.025$); see *Methods* and Fig. S4. The unit cell (Fig. 1a) of the octet-truss resembles a fcc microstructure, such that the octet-truss deforms primarily by axial stretching of the constituent struts *(12)*. Specimens tested in this study had cell sizes ranging $0.15 \text{ mm} \leq \ell \leq 0.75 \text{ mm}$ and relative density $\bar{\rho}$ (viz. the density ratio between the metamaterial and its parent material) in the range $0.01 \leq \bar{\rho} \leq 0.10$. Thus, the largest strut diameter $2r_0 = 65$ μm in the specimens tests is much smaller than the transition flaw size of the TMPTA (SI and Table S1) implying that tensile fracture of the struts in the octet-truss occurs in a purely brittle manner with no associated inelastic deformation. The small cell size compared to specimen sizes implies that no microstructural features are visible via macroscale observations of these specimens (Fig. 1a), and thus they resemble structure-free continua for which there exists a large and well-developed field of fracture mechanics *(18)*. A crack within a periodic truss metamaterial is a continuous array of broken nodal connections with the sharpest possible crack defined by a single missing layer of nodes: removal of one layer of nodes is equivalent to removing one-unit cell in the octet-truss as the remaining struts of the broken unit cell can carry no load (Fig. S5). Here we consider such a sharp square crack of fixed size $2a = 6$ mm embedded centrally in the specimen (Fig. 1a) with the crack corresponding to a missing layer of unit cells; see Figs. 1b-1c and Movie S1. The global Cartesian co-ordinate system $(x, y, z)$ is co-incident with the $[110], [\bar{1}10]$ and $[001]$ directions, respectively of the fcc microstructure, with the crack fronts also co-linear with the $x$ and $y-$directions (Fig. 1a).

As a first step to characterize fracture, we conducted uniaxial tensile tests for loading in the $z-$direction combined with an in-situ XCT observation protocol (Fig. 1d, *Methods*) to monitor failure in the vicinity of the embedded crack. For a macroscopic tensile load $P$ applied perpendicular to the crack plane (Fig. 1a), the mode-I stress intensity factor $K_I$ is given by

$$K_I = Y_I\left(\frac{a}{W}, \frac{E_x}{G_{xy}}, \nu_{xy}\right)\frac{P}{4(WB)}\sqrt{\pi a}, \qquad (1)$$

where $Y_I$ is a calibration factor that depends on specimen geometry parametrised by $a/W$ and ratios of the effective elastic constants (viz. Young's modulus $E_x$ to shear modulus $G_{xy}$ and Poisson's ratio $\nu_{xy}$) of the octet-truss with cubic symmetry; see SI for details of the calibration procedure for $Y_I$. The fracture toughness $K_{IC}$ is then defined as the value of $K_I$ at the failure load $P = P_f$ and we fixed $a/W = 0.2$ for all measurements in line with most continuum fracture testing protocols (continuum mechanics specifies that $K_{IC}$ calculated from a measured $P_f$ is independent of crack size $2a$ when $a/W$ is fixed). The load $P$ versus displacement $u$ for specimens with and without a crack are shown Fig. 1e ($\bar{\rho} = 0.08, a/\ell = 10$). XCT images in Fig. 1f illustrates that the deviation in response of the cracked specimen from its uncracked counterpart is associated with the onset of tensile fracture of crack front struts. The failure load $P_f$ is therefore defined as the load $P$ at which $dP/du$ first reduces to 95% of its initial value. We report measurements on specimens with cracks spanning over 8 to 40 unit cells, i.e. $4 \leq a/\ell \leq 20$, where $a/\ell$ (which quantifies the number of unit cells over the crack flank; see Fig. S6 and SI) is varied by manufacturing specimens of varying cell sizes $\ell$ keeping the overall specimen and crack sizes fixed (Fig. 1c). Dimensional analysis *(16,17,19)* dictates that $\bar{K}_{IC} \equiv K_{IC}/(\sigma_f \sqrt{\ell})$ is then solely a function of $\bar{\rho}$ and $\varepsilon_f$ with the effect of $\ell$ normalized out (Fig. 1g). At high values of $\bar{\rho}$, the failure mode is tensile fracture of struts at the crack front (Fig. 1f and Movie S2), and $K_{IC}$ scales linearly with $\bar{\rho}$, as is well-established *(25,26)* for the stretch-dominated octet-truss lattice. However, the scaling relation changes to $\bar{\rho}^2$ at lower densities, as the failure mode switches to elastic buckling of struts at the crack front (Fig. 1h). Buckling of struts in an octet-truss subjected to tensile loading is well-known *(13)*, but the presence of the crack exacerbates this failure mode in the vicinity of the crack. Although elastic buckling of struts does not necessarily result in strut fracture, it limits structural integrity. This is evidenced by the loss of load carrying capacity and consequent nonlinearity of the $P$ versus $u$ response (Fig. S7) similar to when struts fracture under tensile stresses. Hence, the deviation from linearity is used to define a toughness. More intriguing is the dependence of $\bar{K}_{IC}$ on the number of unit cells in the crack as parameterized via $a/\ell$. The apparent toughness increases with increasing $a/\ell$ more prominently at lower relative densities, where toughness is set by strut elastic buckling. Such dependence of toughness on $a/\ell$ is excluded in continuum fracture mechanics and suggests that $K_{IC}$ is inadequate to characterize the crack-based failure modes of octet-truss lattices at-least for cracks spanning less than 20 unit cells ($a/\ell < 10$); for $a/\ell \geq 10$, the dependence on $a/\ell$ becomes negligible in these uniaxial tensile experiments (Fig. 1g).

The unexpected $a/\ell$ dependence of $\overline{K}_{IC}$ suggests a fundamental lack of understanding of the parameters that govern failure at crack fronts in metamaterials. To investigate further we performed multi-axial loading measurements by applying a combination of axial load $P$ in the $z-$direction and in-plane equi-biaxial loads $Q$ in both the $x$ and $y-$directions (Fig. 2a, S8, and SI). Proportional loading experiments were conducted for tensile loading $P$ while keeping the triaxiality $\lambda \equiv Q/P$ constant. The stress intensity factor $K_I$ is independent of $\lambda$, and thus continuum elastic fracture mechanics requires the measured toughness to be independent of $\lambda$. Measurements of $\overline{K}_{IC}$ as a function of $\lambda$ for two choices of $\bar{\rho}$ and $a/\ell$ to span the tensile fracture and elastic buckling failure modes (Fig. 2b) show a strong dependence of $\overline{K}_{IC}$ on both $\lambda$ and $a/\ell$. This further confirms that $\overline{K}_{IC}$ is an insufficient parameter to characterize failure of these metamaterials.

To help develop a physical understanding of the observations, we performed finite element (FE) simulations for both uniaxial and multi-axial loading cases, where every strut in the specimen is modelled (Fig. 2c, *Methods*). Typical calculations had in excess of 20 billion degrees of freedom. Predictions of $\overline{K}_{IC}$ (Figs. 1g and 2b) and the corresponding failure modes (Figs. 2d-2g) illustrate the fidelity of the model in capturing both tensile and elastic buckling failure modes. While the measured $\overline{K}_{IC}$ is highly repeatable, the precise location of struts in elastic buckling varies from crack front (Fig. 2h) to crack flank (Fig. 2i and Movie S3) in nominally equivalent experiments. The difference in failure loads for the crack front and flank buckling modes (Figs. 2e and 2f) is predicted as less than 2% for the $\bar{\rho} = 0.03$ and $a/\ell = 4$ specimens, and this holds for all cases considered here. Thus, it is the inevitable small imperfections in the as-manufactured specimens that sets the precise buckling failure mode, thereby rationalizing the variability of buckling modes observed in the experiments (Figs. 1h, 2h and 2i). The fidelity of the FE predictions where every strut within the specimen is discretely modelled allows us to use these calculations to probe the apparent inadequacy of $K_{IC}$ to characterize fracture.

### Hunt for the source of the crack size and stress triaxiality dependence of $K_{IC}$

Fracture of elastic continua is well-characterized by toughness $K_{IC}$, but here we have demonstrated that toughness for the octet-truss metamaterial depends on both $\lambda$ and $a/\ell$. Under plastic deformations, the stress-triaxiality is known to influence fracture via the so-called T-stress. While plasticity is not operative in our experiments (TMPTA is elastic-brittle; see Fig. S4), the strong influence of stress triaxiality as parameterized by $\lambda$ suggests that T-stresses might surprisingly influence the elastic fracture of the octet-truss.

An asymptotic analysis of an elastic continuum with cubic symmetry dictates that the stresses $\sigma_{ij}$ around the crack tip, measured in the crack front co-ordinate system $e_i$ (Fig. 2g), scale as

$$\tilde{\sigma}_{ij} \equiv \frac{\sigma_{ij}}{\sigma_f} = \bar{K}_I \left[ \bar{r}^{-\frac{1}{2}} f_{ij}^I + \bar{T}_{11} \delta_{1i} \delta_{1j} + \bar{T}_{33} \delta_{3i} \delta_{3j} + \bar{T}_{13} \delta_{1i} \delta_{3j} + \mathcal{O}(\bar{r}^{\frac{1}{2}}) + \mathcal{O}(\bar{r}) + \ldots \right], \quad (2)$$

where $(r, \theta)$ are the in-plane polar coordinates in a plane normal to the crack front, $e_1$ the direction formed by the intersection of the plane normal to the crack front and the crack plane, and $e_3$ a direction tangential to the crack front within the crack plane. In (2) the normalized quantities are $\bar{r} \equiv r/\ell$, $\bar{K}_I \equiv K_I/(\sigma_f \sqrt{\ell})$ and $\bar{T}_{ij} \equiv T_{ij}\sqrt{\ell}/K_I$, with the non-dimensional function $f_{ij}^I(\theta, E_x/G_{xy}, \nu_{xy})$ governing the angular dependence of the singular stress terms, while $T_{ij}$ are the so-called T-stress components and $\delta_{ij}$ is the Kronecker delta. The discreteness of the lattice (i.e. a unit cell of size $\ell$) implies that a mathematically sharp crack front cannot be defined. It is thus conceivable that fracture is governed not only by the most singular term in the expansion (2), which scales with $K_I$, but also the T-stress terms that are finite in the vicinity of $r = \ell$. To test this hypothesis in a simplistic setting, we considered a one unit-cell thick slice through the octet-truss specimen (Fig. 3a) under plane-strain conditions and investigated the fields by performing an asymptotic analysis *(19)*. In particular, we considered a semi-infinite crack and imposed remote displacement boundary conditions consistent with only the $K_I$ and $T_{ij}$ terms in the expansion (2); see SI. In this plane-strain setting, with deformation restricted to the $e_1 - e_2$ plane and the crack along the $e_1$-direction, $T_{13} = 0$ and it suffices to specify the T-stress via $T = T_{11}$ as the plane-strain condition then requires $T_{33} = \nu_{xy} T$. Thus, loading is uniquely specified by the combination $(\bar{K}_I, \bar{T} \equiv T\sqrt{\ell}/K_I)$.

Predictions (Fig. 3b) of $\bar{K}_{IC}$ in this plane-strain setting for $\bar{T}$ in the range $-1 \leq \bar{T} < 1$ are similar to the embedded crack 3D case, with fracture for high values of $\bar{\rho}$ set by tensile fracture of the struts and the failure mode transitioning to elastic strut buckling at lower $\bar{\rho}$. These generalized 2D results are not only qualitatively consistent with the 3D measurements but also quantitatively very similar (observe the axes scales in Figs. 1g and 3b). Moreover, the strong effect of $\bar{T}$ on the predictions is analogous to that of $\lambda$ in the sense that $\bar{K}_{IC}$ decreases with increasing $|\bar{T}|$ for a fixed $\bar{\rho}$. To understand this influence of $\bar{T}$, we performed a stress analysis and show the axial stress distributions (Fig. 3c) in struts around the crack tip for an applied $K_I/(E_s\sqrt{\ell}) = 0.01$ and three choices of $\bar{T}$. The axial stresses $\sigma_a$ are normalized as $\bar{\sigma}_a \equiv \sigma_a/(E_s\bar{\rho})$ so that $\bar{\sigma}_a$ in Fig. 3c is independent of $\bar{\rho}$. Both tensile and compressive axial loads develop in struts around the crack tip, and hence failure can be governed by either strut tensile fracture or elastic buckling. The stress analysis illustrates that both maximum tensile and compressive stresses $\bar{\sigma}_a$ increase with increasing $|\bar{T}|$. This strong influence of $\bar{T}$ is a consequence of the discreteness of the micro-structure which means that the 2nd term (non-singular) in the asymptotic expansion (2) cannot be neglected, with $\bar{T}$ strongly influencing the fracture mechanisms.

*A fracture mechanism map*

The asymptotic analysis suggests that T-stress effects are essential in developing a fracture criterion for the octet-truss metamaterial. But can the T-stress explain both the observed triaxiality $\lambda$ and $a/\ell$ dependence of $\bar{K}_{IC}$? The calibration analysis in (1) is extended to determine the T-stress for a mathematically sharp crack embedded in a 3D anisotropic elastic continuum (SI, Fig. S9). Using the crack-front co-ordinate system $e_i$ (Fig. 3a) and labelling $T = T_{11}$, we define $\bar{T} \equiv \hat{T}\sqrt{\ell/a}$ where $\hat{T} = T\sqrt{a}/K_I$ is a calibration constant for the T-stress commonly known as the biaxiality ratio. The calibration of $\hat{T}$ for our cubic specimen with an embedded square crack gives (SI)

$$\bar{T} \approx \frac{\sqrt{\pi}}{2.2}(\lambda - 1)\sqrt{\frac{\ell}{a}}, \tag{3}$$

and thus $\bar{T}$ is a function of both $\lambda$ and $a/\ell$. The data (measurements and FE predictions) of Fig. 2b are replotted (Fig. 3d) using a $x$−axis of $\bar{T}$ given by (3) rather than $\lambda$. For a given $\bar{\rho}$, the data for both values of $a/\ell$ collapse onto a unique curve demonstrating that both the observed $a/\ell$ and $\lambda$ dependence of $\bar{K}_{IC}$ is in fact a T-stress effect. Thus, varying $\lambda$ is equivalent to changing $a/\ell$ in terms of the influence on the effective toughness. These findings can be summarized in the form of a fracture mechanism map (Fig. 3e), where the $y$−axis is the material property $\bar{\rho}/\varepsilon_f$ while $x$−axis is the structural parameter $\bar{T}$ set by a combination of $a/\ell$ and $\lambda$ (3). These structural and material parameters set the operating point on the map from which we can read-off the normalized toughness $\bar{K}_{IC}/\varepsilon_f$: we emphasize that this map is valid for an octet-truss metamaterial made from any elastic parent material irrespective of its failure strain $\varepsilon_f$ (see SI for a discussion on the normalizations used in Fig. 3e so as to make it valid for all $\varepsilon_f$). The failure regimes (strut elastic buckling and tensile strut fracture) are shaded, and we observe that the transition between the failure modes is strongly dependent on $\bar{T}$. This is especially true for $\bar{T} < 0$, where the relative density to transition from an elastic buckling to tensile fracture failure mode increases with decreasing $\bar{T}$ as compressive T-stresses enhance the propensity for strut buckling. This procedure to generate fracture mechanism maps is general and independent of topology. Further examples of such maps are provided in Figs. 3f-3h: (i) elastically isotropic and stretch-dominated compound truss with by volume 40% simple cubic (SC) and 60% body-centered cubic (BCC) and labelled 40SC/60BCC (or iso-SC/BCC); (ii) the elastic cubic and stretch-dominated compound truss with by volume 60% simple cubic (SC) and 40% body-centered cubic (BCC) and labelled 60SC/40BCC (or aniso-SC/BCC) and (iii) the bend-dominated chiral gyroid topology (in all cases the toughness is for mode-I loading in the $e_1 - e_2$ plane with the co-ordinate system $e_i$ for each topology defined in Figs. 3f-3h). Qualitative features of the fracture mechanism map remain unaltered for this wide range of truss topologies ranging from stretch to bending-dominated as well as elastically isotropic and anisotropic with all topologies displaying a strong dependence of the toughness on $\bar{T}$ which in most cases is significantly stronger than that observed for the octet-truss. The one difference that emerges is that elastic strut buckling dominated fracture mode is not operative for the bend-dominated gyroid topology.

The maps Figs. 3e-3h can be combined to construct a topology selection map (Fig. 3i) where we mark the topology that maximizes $\bar{K}_{IC}/\varepsilon_f$ for a given combination of $(\bar{\rho}/\varepsilon_f, \bar{T})$. The aniso-SC/BCC and octet truss topologies dominate the map with the iso-SC/BCC topology only optimal in a narrow range. By contrast, as would be anticipated the bend-dominated gyroid always has a lower toughness compared to these stretch-dominated topologies and does not appear on the topology selection map. We emphasize that these fracture mechanism maps (Figs. 3e-3f) are independent of the crack or specimen shape, size and loading configuration. The geometrical and loading dependence of fracture is captured by the calibration parameters of $T$ and $K_I$ in any given structural situation and examples of the associated calibration constants $\hat{T}$ and $Y_I$ for structures made from the octet-truss and subjected to a wide range of loadings are provided in Fig. S10.

*A design protocol*

Our fracture mechanism map uncovers the role of fracture toughness for the failure of architected metamaterials under arbitrary loadings. In contrast to continuum fracture mechanics where fracture toughness is a single parameter that governs failure in an elastic material, Fig. 3e reveals that the failure of a metamaterial is governed by two parameters which link loading and specimen geometry ($\bar{T}$) and micro-structural/parent material parameters ($\bar{\rho}/\varepsilon_f$) to the apparent fracture toughness of the metamaterial. Crucially, this map points to a general design and evaluation protocol to assess the damage tolerance of architected metamaterials for arbitrary structural loadings. To illustrate the use of this map, here we discuss the scheme to evaluate the failure load $P = P_f$ of a cracked octet-truss metamaterial beam under 4-point bending: the beam has an embedded square crack of side $2a$ symmetric with respect to the neutral axis (Fig. 4a). The steps in the scheme are summarized as follows. First, perform continuum elastic calculations (Fig. 4b), analogous to traditional K-calibration calculations (SI, Fig. S11), to provide the calibration factors $Y_I = 1.34$ and $\hat{T} = 1.13$ of the specimen. These calibration factors combined with the metamaterial microstructural parameters ($a/\ell, \bar{\rho}$) and parent material property $\varepsilon_f$, enable us to locate $\bar{T}$ and $\bar{\rho}/\varepsilon_f$ on the cross-plot (Fig. 4c) of the fracture mechanism map (Fig. 3e) to read-off $\bar{K}_{IC}/\varepsilon_f$. Then given $\bar{K}_{IC}$ the normalized failure load $\bar{P}_f \equiv sP_f/(8\sigma_f W^3)$ follows directly from the calibrated value of $Y_I$. Predictions (Fig. 4d) of the normalized failure load of the $\bar{\rho} = 0.08$ octet-truss are plotted for two different parent materials with $\varepsilon_f = 0.025$ and $0.1$ for crack sizes varying from 8-unit cells ($a/\ell = 4$) upto 40. Fracture is dominated by tensile strut failure for $\varepsilon_f = 0.025$ but set by elastic strut buckling when $\varepsilon_f = 0.1$. Predictions wherein failure is assumed to be dominated by the K-field (i.e. assuming that $\bar{T} = 0$ or equivalently ignoring the T-stress effects) are also included in Fig. 4d and overpredict the failure loads. The overprediction increases with decreasing $a/\ell$ (~ 45 % and ~ 30% at $a/\ell = 4$ for the $\varepsilon_f = 0.10$ and 0.025 cases, respectively) but even for specimens with cracks spanning 40 unit cells ($a/\ell = 20$) the $K$−dominated assumption results in a 10% overprediction. Nearly all existing experimental studies use bend specimens *(16)* and have cracks spanning no more than 10 cells – we thus anticipate that these studies have significantly overestimated toughness.

The protocol can also be used to determine the topology that maximizes structural strength. For example, while Fig. 4d gives the failure load for a $\bar{\rho} = 0.08$ octet-truss beam in bending we may alternatively wish to determine the topology (Fig. 4e) that maximizes the 4-point bending strength. Combining the calibration factors $(Y_I, \hat{T})$ (Fig. 4f) for the four different topologies with the fracture mechanism maps (Figs. 3e-3f) gives the normalized failure loads $\bar{P}_f$ for the cracked beam ($a/\ell = 16$) in 4-point bending. A combination $\bar{T}$ dependence of the toughness and the elastic anisotropy of the metamaterials implies that the aniso-SC/BCC is optimum for a beam made from a low failure strain parent material with $\varepsilon_f = 0.025$ while the octet-truss topology maximizes $\bar{P}_f$ at the higher value of $\varepsilon_f = 0.1$ (Fig. 4g). This general protocol can be employed for any arbitrary loading with the results in Fig. 4 further emphasizing that toughness alone is insufficient not only to characterize failure of metamaterial structures but also inadequate to select an optimal metamaterial topology in a given application.

*__The reversal of plane-strain and plane-stress toughness__*

A thick specimen with a through-thickness crack is conventionally used to evaluate the thickness-independent plane-strain fracture toughness *(23)*. This measured value is typically quoted as a material property based on the fact that the plane-strain toughness is lower than the plane-stress toughness (thin specimens) in standard continuum solids *(27)*. (Fracture initiates within the specimen where plane-strain conditions prevail rather than on the specimen surfaces where the conditions resemble plane-stress, and the toughness is higher.) To-date all fracture tests on metamaterials have used through-thickness cracks *(14-17)*, but it remains unclear whether the underlying assumption of the plane-stress toughness exceeding the plane-strain toughness remains valid in these materials. If not, these tests have not measured the plane-strain toughness.

To investigate the use of through-thickness cracked specimens for the measurements of toughness, we considered center-cracked tension (CCT) specimens (Fig. 5a) of in-plane dimensions $2W = 2H = 35$ mm. These specimens can be viewed as a slice through the mid-section of the 3D embedded crack specimen (Fig. 3a). The thickness $2B$ of the specimens (which is in the $[\bar{1}10]$ direction) was varied over $\ell \leq 2B \leq 100\ell$ to examine the plane-stress and plane-strain limits. Toughness $K_Q$ (toughness inferred in this way is denoted as $K_Q$, as it is unclear whether it equals the plane-strain limit) is calculated using (1) with $Y_I$ recalibrated for the CCT specimens (SI). Measurements of $\bar{K}_Q$ ($\bar{K}_Q \equiv K_Q/(\sigma_f \sqrt{\ell})$ ) versus $\bar{\rho}$ for two choices of $a/\ell$ (Fig. 5b) reveal a scaling behavior very similar to the embedded crack specimens (Fig. 1g) although the sensitivity to $a/\ell$ is milder. The corresponding FE predictions show excellent agreement with measurements (Fig. 5b), and the XCT observations (Figs. 5c to 5h), which reveal that tensile strut fracture and strut elastic buckling, are again the two operative failure modes. Intriguingly $\bar{K}_Q$ is insensitive to thickness over $\ell \leq 2B \leq 100\ell$. This surprising observation is complemented by the fact that $\bar{K}_Q$ is lower than $\bar{K}_{IC}$ as measured from the embedded crack specimen for a given $a/\ell$ (results from Fig. 1g reproduced in Fig. 4b). This is all the more puzzling since $\bar{K}_Q$, which is the toughness of the $2B = \ell$ specimen, can be interpreted as the plane-stress toughness, while the embedded crack $\bar{K}_{IC}$ measurements have no specimen surface effects and hence representative of a plane-strain toughness. Thus, unlike conventional continuum materials the plane-stress toughness is lower than the plane-strain toughness for the octet-truss metamaterial. To understand this reversal, observe that failure even for the $2B = 100\ell$ specimen is dominated by failure of surface struts (Figs. 5g and 5h and Movie S4). FE predictions of the tensile and compressive axial stresses in the struts across the specimen thickness for the $2B = 100\ell$ specimen (Fig. 5a) confirm that both the compressive and tensile strut stresses are a maximum at the surface. These high surface stresses result from the surface nodes of the metamaterial being "unbalanced"; i.e. struts are absent on one side causing higher stresses to build-up in surface struts resulting in their premature failure. In a specimen with a through-thickness crack, such free-surfaces are always present irrespective of the specimen thickness, and thus fracture initiation is always surface-dominated. Therefore, two distinct characteristic toughnesses exist for metamaterials: one for through-thickness or surface flaws and another for embedded flaws. The equivalent of a plane-strain toughness can only be measured via an embedded crack specimen that eliminates surface effects.

*Outlook*

Existing fracture analyses of elastic metamaterials *(16,17, 19-22)* have a-priori (and without subsequent experimental validation) assumed that the stress-intensity factor is sufficient to characterize fracture, and then used FE calculations to develop scaling laws for the toughness. Such scalings can also be extracted from our measurements and analyses. For example, in the absence of a T-stress (i.e. $\bar{T} = 0$), our 3D embedded crack data suggests that the normalized toughness $\bar{K}_{IC}$ scales as

$$\bar{K}_{IC} = \begin{cases} \alpha_0 \bar{\rho} & \text{for } \bar{\rho} \geq \dfrac{\alpha_0}{\beta_0} \varepsilon_f \\ \dfrac{\beta_0}{\varepsilon_f} \bar{\rho}^2 & \text{otherwise,} \end{cases} \quad (4)$$

where $\alpha_0 = 0.31$ and $\beta_0 = 0.33$, with tensile strut fracture being the operative failure mode for $\bar{\rho} \geq (\alpha_0/\beta_0)\varepsilon_f$. Here we have shown that such scaling laws cannot be used to predict fracture in components made from the octet-truss metamaterial, as $\alpha_0$ and $\beta_0$ are dependent on the number of unit cells over the crack flank as parameterized by $a/\ell$ and the imposed stress-triaxiality $\lambda$. The effect of both $a/\ell$ and $\lambda$ is captured via a single parameter $\bar{T}$ that is related to the so-called T-stress in the asymptotic expansion of the crack-tip fields. Unlike in 2D (two-dimensional) planar networks *(26)* the T-stress effect is shown to have a very large effect for a wide-range of 3D metamaterials with stress-triaxiality effects amplifying the role of T-stresses in governing fracture and these T-stress effects. The T-stress is a non-singular term which is neglected in continuum elastic fracture as fracture is a highly localized process that is dominated by singular stresses, which scale with the stress-intensity factor $K_I$. However, the discreteness of the octet-truss metamaterial implies that a mathematically sharp crack is not present, and hence a strut that undergoes either tensile fracture or elastic buckling failure is at finite distance from the ill-defined crack front. In such a scenario, the non-singular T-stress terms become of similar order to the $K_I$ terms, creating the need for an enrichment of conventional elastic fracture mechanics to describe fracture of elastic metamaterials irrespective of their microstructure. Another consequence of the discreteness of the microstructure is the inadequacy of conventional fracture testing protocols to measure material properties that describe fracture of 3D metamaterials. Specifically, contrary to all reported toughness measurements *(16,17)*, the equivalent of a plane-strain toughness in 3D metamaterials can only be measured via embedded crack specimens that eliminates specimen surface effects.

T-stress effects have been reported in inelastic materials *(28-32)* when the size of the damage process zone is comparable to the crack size. No such finite process zone exists in our case as not only seen in the X-ray micrographs (Fig. 1f) but also clear from the fact that the transition flaw size of the TMPTA is nearly 20 times the largest strut diameter of the truss-based metamaterials we have investigated. Rather here we have demonstrated that T-stress effects play a significant role in elastic metamaterials due to the discreteness of the microstructure: in the limit of a microstructure free continuum metamaterial with $\ell/a \to 0$, $\bar{T} \to 0$ (3) and the effect of T-stresses vanish. Such discreteness effects are also expected to play a role in atomistic simulations of brittle fracture wherein too breakdown of conventional fracture mechanics *(33, 34)* has been reported with no unique value of the stress intensity factor characterizing fracture. The general framework that we have introduced here for predicting failure of brittle metamaterials forms the basis of understanding the brittle fracture of such discrete material systems of which metamaterials is an important class but also includes many other important engineering materials such as fiber composites. The framework hinges on the concept of a fracture mechanism map and here we have shown this idea to hold for a broad range of truss-based metamaterials with range of anisotropies and deformation mechanisms. Such maps can readily be used by designers to evaluate failure in different applications at negligible computational cost compared to modelling the microstructural details of the metamaterial.

*Figures:*

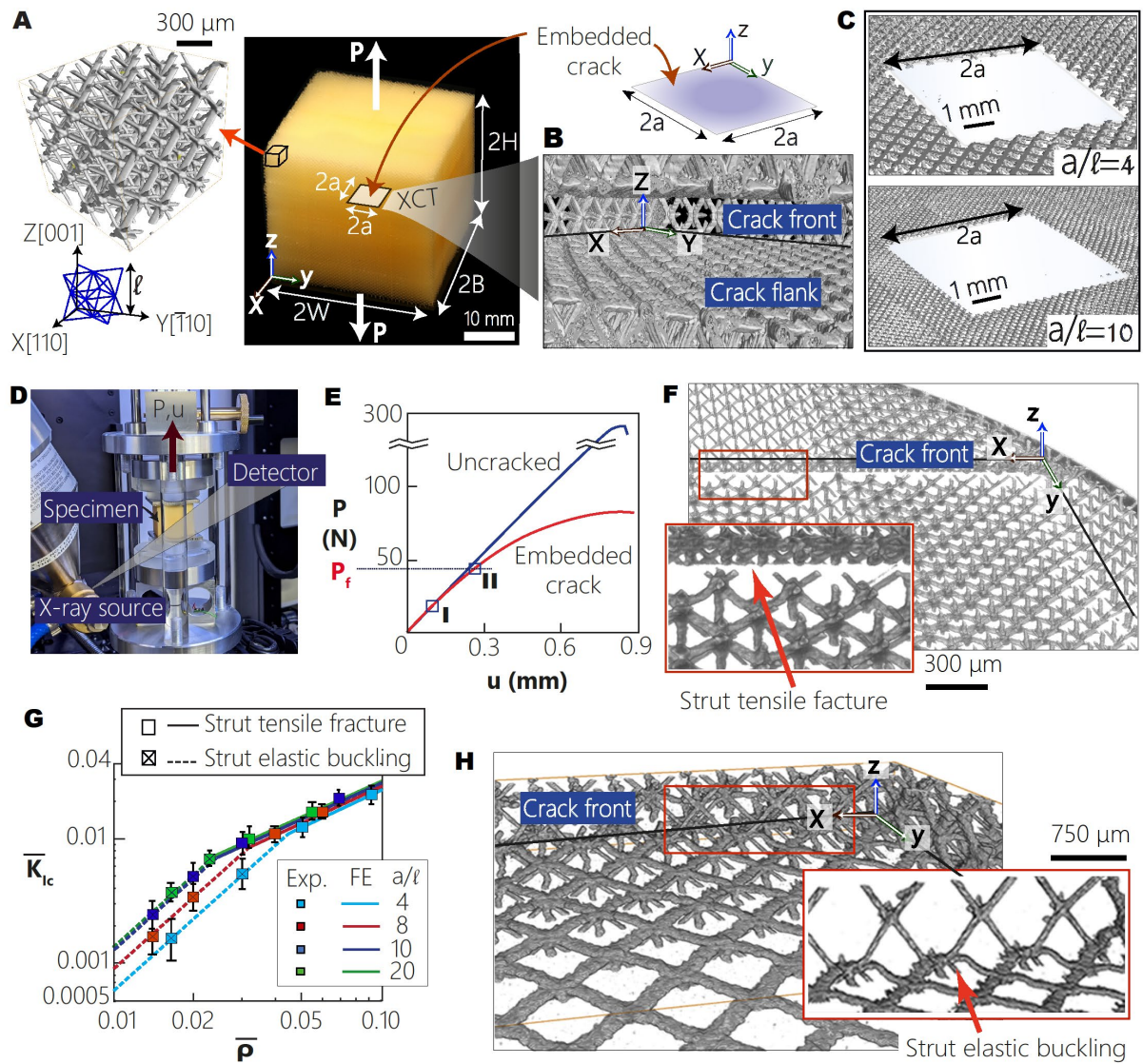

**Figure 1: Failure under uniaxial tensile loading.** (A) Optical image of the cubic octet-truss specimen with an embedded square crack of side $2a$. The insets show XCT images of the octet-truss microstructure and unit cell. XCT images of (B) the crack front/flank at loading-stage I in (E), and (C) cracks of fixed size $2a$ but with varying cell sizes $\ell$ as parametrized by $a/\ell$. (D) Tensile loading setup, with in-situ XCT imaging. (E) Tensile load $P$ versus displacement $u$ response of the $\bar{\rho} = 0.08$ uncracked and cracked ($a/\ell = 10$) specimens. (F) The crack front/flanks at $K_{IC}$ (loading-stage II in (E)). (G) Measured normalized toughness $\bar{K}_{IC}$ versus relative density $\bar{\rho}$ (lines are FE predictions and symbols are measurements with error bars indicating variation over 5 test samples). (H) XCT image for the $\bar{\rho} = 0.03$ and $a/\ell = 4$ specimen when failure is set by elastic buckling of crack front struts (inset shows a magnified view of the buckled struts).

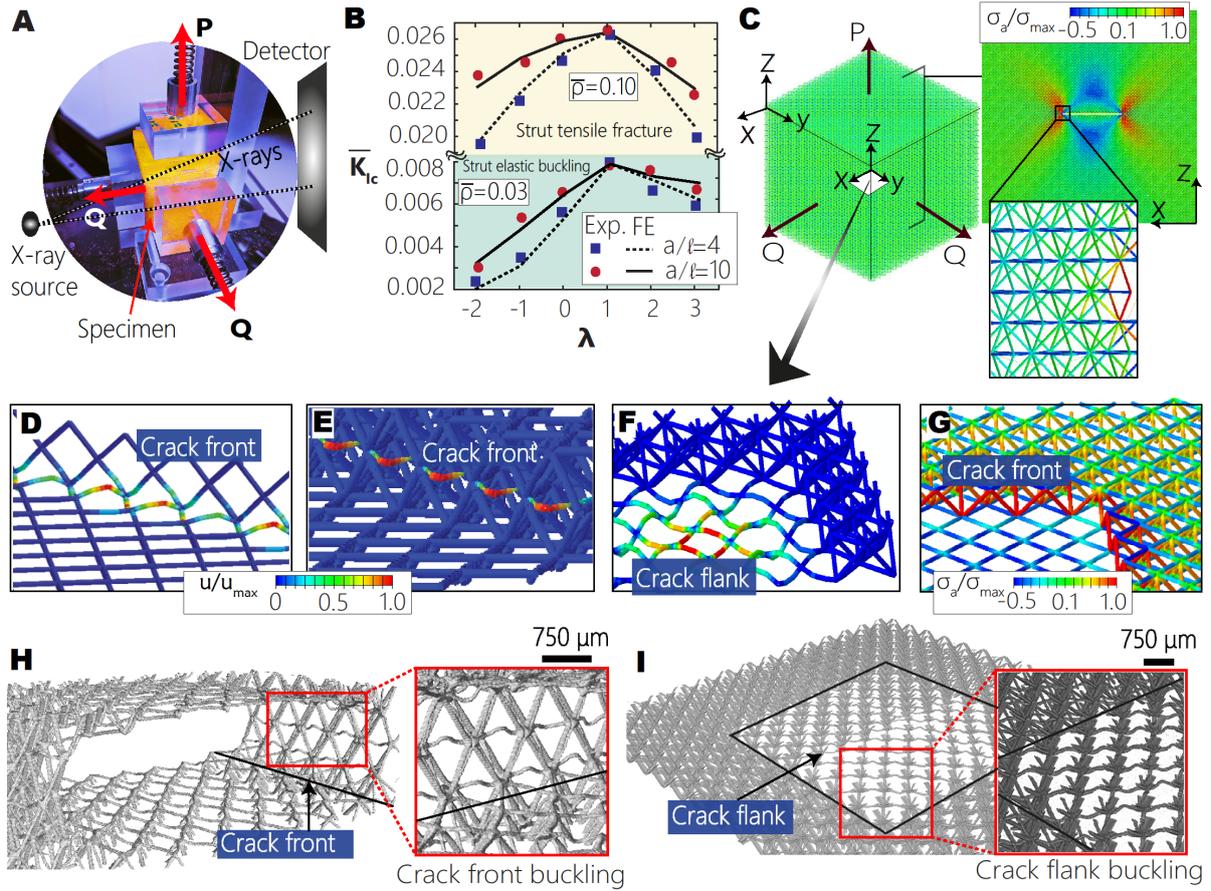

**Figure 2: Failure under multiaxial loading.** (A) Multi-axial loading setup with in-situ XCT imaging. (B) Summary of the measurements and FE predictions of $\overline{K}_{IC}$ as a function of the load triaxiality $\lambda \equiv Q/P$ for specimens that fail by strut fracture ($\bar{\rho} = 0.1$) and elastic strut buckling ($\bar{\rho} = 0.03$). Results are shown for two values of $a/\ell$ in both cases. (C) FE modelling of the specimens, with insets showing each purely elastic strut modelled individually shaded by contours of the normalized axial stress $\sigma_a/\sigma_{max}$ ($\sigma_{max}$ is the maximum axial strut stress in the entire specimen). (D-G) FE predictions of the failure modes for loading with $\lambda = 1$. Failure set by elastic buckling of (D-E) crack front and (F) crack flank struts of the $\bar{\rho} = 0.03$ and $a/\ell = 4$ specimen. The predicted $\overline{K}_{IC}$ differs by 1% between (D) and (E), and 2 % between (E) and (F). The struts are shaded by magnitude of normalized displacement $u/u_{max}$. (G) Distribution of normalized axial stresses $\sigma_a/\sigma_{max}$ for the $a/\ell = 10$ and $\bar{\rho} = 0.10$ specimen with crack-tip struts predicted to undergo tensile fracture at $K_{IC}$. XCT images for the $\bar{\rho} = 0.03$ and $a/\ell = 4$ specimen for loading with $\lambda = 1$ when failure is set by elastic buckling of (H) crack front struts and (I) crack flank struts (insets shows a magnified view of the buckled struts). The measured $K_{IC}$ between (H) and (I) differs by only 2%.

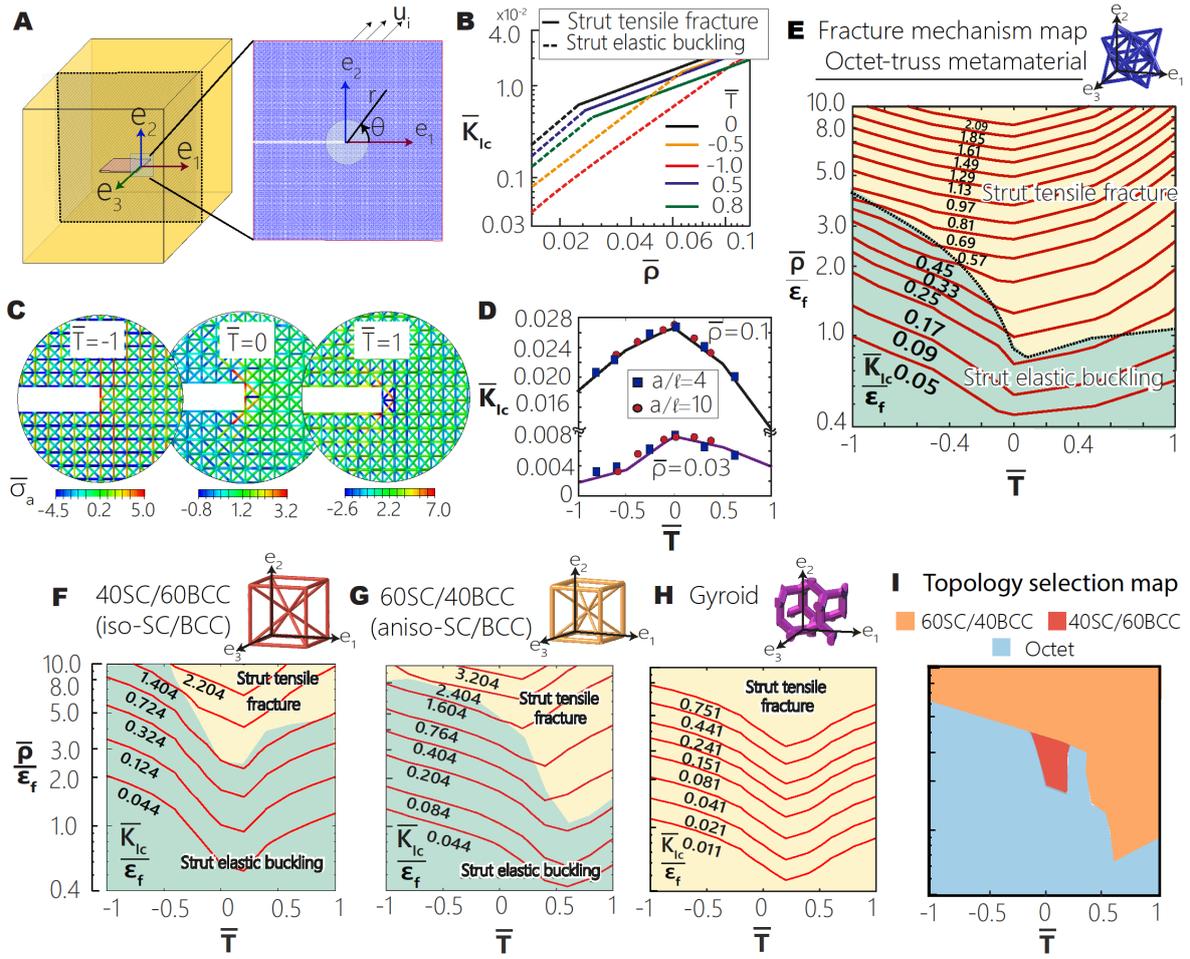

**Figure 3: Fracture mechanism maps.** (A) Sketch illustrating the crack front co-ordinate system and a 2D slice used in the asymptotic analysis. (B) FE predictions showing the sensitivity of the asymptotic predictions of $\overline{K}_{IC}$ to $\overline{T}$ over a range of relative densities $\overline{\rho}$. (C) FE predictions of normalized axial stress $\overline{\sigma}_a \equiv \sigma_a/(E_s\overline{\rho})$ around the crack tip for an applied $K_I/(E_s\sqrt{\ell}) = 0.01$ and three choices of the normalized T-stress $\overline{T}$. (D) FE predictions and measurements from Fig. 2b replotted as a function of $\overline{T}$ to illustrate that $\overline{\rho}$ and $\overline{T}$ set $\overline{K}_{IC}$, with the effect of $a/\ell$ and $\lambda$ both captured within $\overline{T}$. (E) Fracture mechanism map of the octet-truss metamaterial, with axes of normalized T-stress $\overline{T}$ and $\overline{\rho}/\varepsilon_f$ and contours of $\overline{K}_{IC}/\varepsilon_f$. The strut tensile fracture and strut elastic buckling failure regimes are shaded. Corresponding fracture mechanism maps of the stretch-dominated (F) isotropic compound simple cubic (SC) and body centered-cubic (BCC) truss (G) anisotropic compound SC/BCC truss (H) bend-dominated gyroid and (I) a topology selection map indicating which topology maximizes $\overline{K}_{IC}/\varepsilon_f$ in different regions of $\overline{\rho}/\varepsilon_f$ versus $\overline{T}$ space.

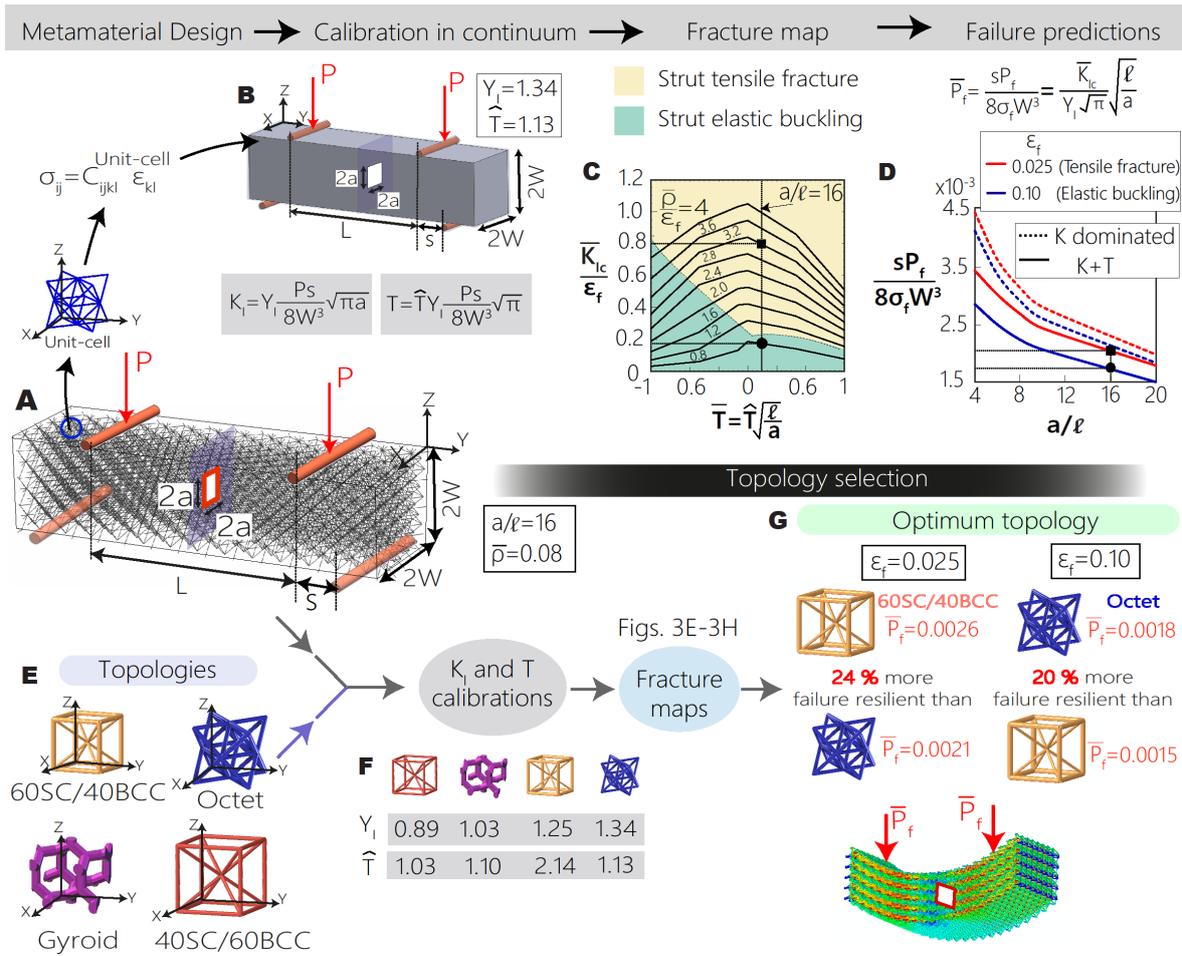

**Figure 4: Design with metamaterials.** (A) Octet-truss beam of aspect ratio $L/W = 20$ with the embedded crack $(a/W = 0.2)$ subjected to four-point bending. (B) Geometry of the continuum anisotropic elastic beam used to determine the calibration factors $Y_I$ and $\hat{T}$ for $K_I$ and $T$, respectively. (C) The cross-plotted fracture map from Fig. 3E. (D) Prediction of the normalized failure load $\bar{P}_f$ of the $\bar{\rho} = 0.08$ octet-truss over a range of crack sizes and two choices of parent materials. The reference prediction for an assumed $\bar{T} = 0$ is also included. The black markers in (C-D) show examples of the prediction of the normalized failure load $P = P_f$ for a crack with $a/\ell = 16$ in a $\bar{\rho} = 0.08$ octet-truss metamaterial beam made from parent materials with $\varepsilon_f = 0.025$ and $0.1$. (E-G) Topology selection for maximizing failure load under four-point bending. (E) The four candidate topologies with their orientations labelled in the global beam co-ordinate system $(X, Y, Z)$. (F) Continuum calibration of the geometric constants $Y_I$ and $\hat{T}$ and (G) description of the optimal topology and improvement over the next best candidate for a $\bar{\rho} = 0.08$ beam made from a parent material with failure strain $\varepsilon_f = 0.025$ and $0.1$.

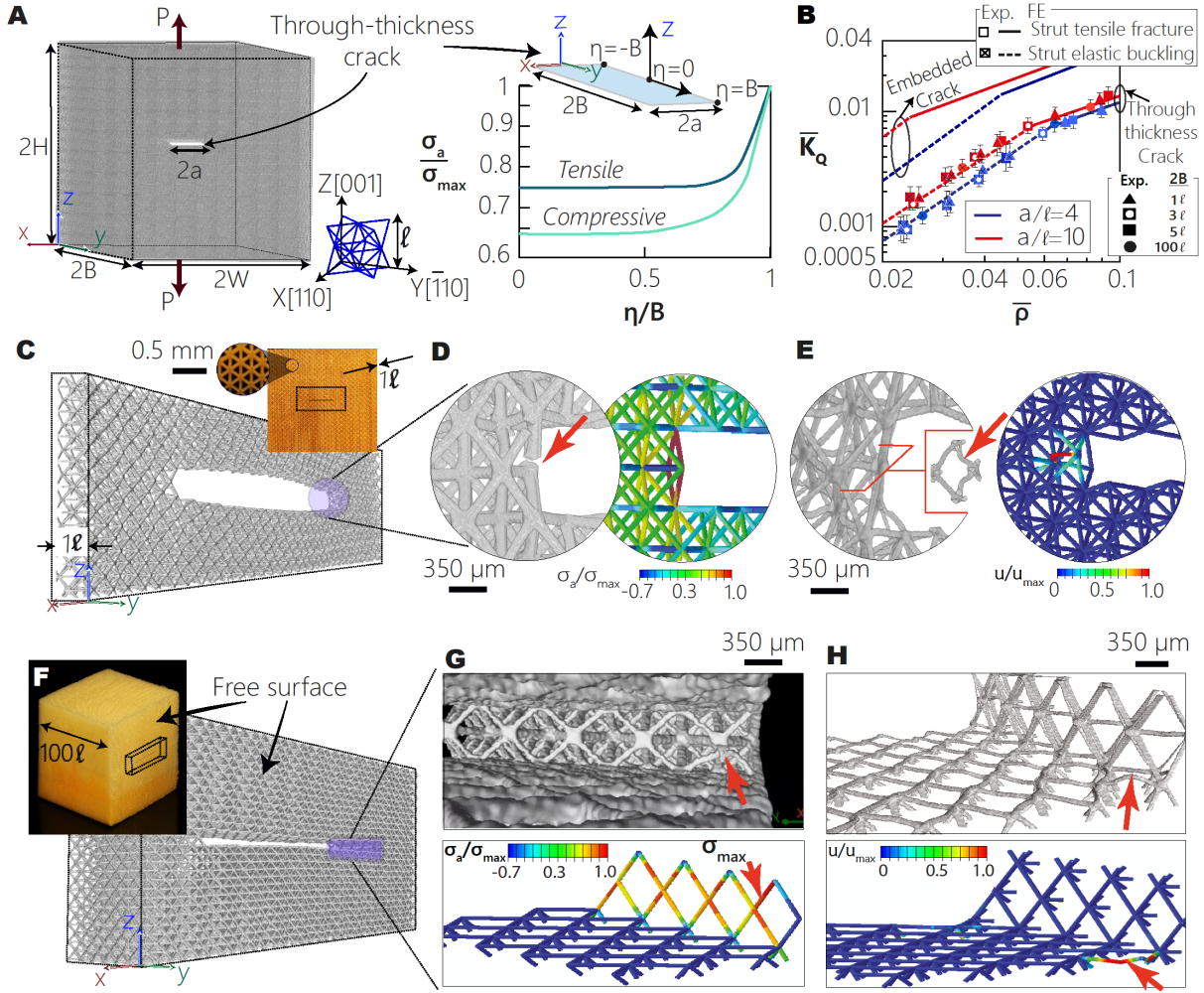

**Figure 5: Through-thickness cracks.** (A) Sketch illustrating tensile loading of a specimen of thickness $2B$ with a through-thickness crack (CCT geometry). FE predictions of the variation of normalized axial strut stresses $\sigma_a/\sigma_{max}$ for struts along the crack front (the maximum tensile and compressive stresses are plotted in each unit cell along the crack front) are included in the inset, which also illustrates the local axis $\eta$ along the crack front with $\eta = 0$ at the mid-plane of the specimen and $\eta = \pm B$ on the specimen free-surfaces. (B) FE predictions and measurements (error bars indicating variation over 5 tests) of $\overline{K}_{IC}$ versus $\bar{\rho}$ for specimens of different thicknesses $2B$. In each case, the results are shown for two values of $a/\ell$, with the embedded crack results reproduced from Fig. 1G. (C) XCT image of a portion of the $2B = \ell$ specimen. Optical image of the specimen along with a magnified view of the microstructure is shown in the inset (scale bar refers to inset). XCT images and FE predictions of the crack tip state at $K_{IC}$ in the $a/\ell = 10$ specimen illustrating the (D) tensile fracture ($\bar{\rho} = 0.10$) and (E) elastic strut buckling ($\bar{\rho} = 0.03$) failure modes. The FE predictions show distributions of the normalised axial stress $\sigma_a/\sigma_{max}$ and the normalised displacements $u/u_{max}$ in (D) and (E), respectively, with the maximum taken over all struts in the specimen. (F) XCT image of a portion of the $2B = 100\ell$ specimen, with $a/\ell = 10$ showing the free surface and 5-unit cells along thickness. Optical image of the specimen is shown in the inset with the dashed cuboid marking the region of the XCT image. (G-H) The corresponding FE predictions and XCT observations show that failure by tensile strut fracture ($\bar{\rho} = 0.10$) and elastic strut buckling ($\bar{\rho} = 0.03$) occurs on the specimen free surfaces.

**Methods**

*Materials:* The octet-truss metamaterial specimens were fabricated using the liquid monomer trimethylolpropane triacrylate (TMPTA; CAS 15628-89-5). The TMPTA was mixed with 1-phenylazo-2-naphthol (Sudan 1; CAS 842-07-9) photo-absorber to adjust the light penetration depth and phenylbis (2, 4, 6-trimethylbenzoyl) phosphine oxide (Irgacure 819) that served as a photo-initiator. The resulting solid material had a tensile failure strength, $\sigma_f \approx 11$ MPa, and the Young's modulus, $E_s \approx 430$ MPa for specimens with dimensions on the order of mm down to the micron scale representative of single struts within the octet-truss specimens (Fig. S4). Nano-indentation tests were also conducted on single struts extracted from the printed specimens. The solid material modulus inferred from these tests varied by less than 5% for indentation carried out over a wide range of orientations thereby confirming the isotropy of the as-printed material. The key advantage of TMPTA is that it has a linear elastic-brittle tensile response and a low strain rate sensitivity (less than 20% change in modulus and fracture strength over three decades of strain-rate).

*Fabrication*: In conventional large-area projection stereolithography (SLA), each layer is printed in a single projection with heat dissipation limited to peripheral cooling. This results in entrapment of heat in the central region, especially for large exposure areas (Fig. S1a). Before heat is fully dissipated, the exposure of next layer starts, and heat accumulates as layer number increases, making the polymerization highly non-uniform within the large printing area. The printed part thus suffers from non-uniform mechanical properties. To overcome this drawback, our fabrication method (Fig. S1b) used a moving spotlight projection SLA technique where curing occurs in sub-sections that are exposed by a continuously moving spotlight (Fig. S1c). There is thus less heat generated compared to exposing the entire cross-section and this results in more uniform polymerization compared to conventional SLA. Additionally, the moving spotlight means that heat is continuously dissipated with minimal accumulation. This unique method allows the production of large-scale parts (100 mm in overall size) with micron scale resolution, low level of imperfections and spatially uniform mechanical properties as validated in our experiments (Fig. S2 and S3).

***Experimental methodology*:** Custom experimental rigs (Fig. 1d and Fig. S8) were designed and fabricated to perform uniaxial (Fig. 1d) and multiaxial loadings (Fig. 2a) with in-situ XCT observations. Specimens were first scanned in the as-manufactured state for quality check and relative densities were then computed using volume analysis available in the commercial software VGStudio Max 3.4. Two different scan protocols were followed- quick scan (15 min acquisition time) and long scan (50 min) for in-situ observations and analysis, respectively (details in Section C of SI). A typical test involved 10-15 quick scans acquired at regular intervals along measured load-displacement curve (Fig. 1e). The crack tip and crack flank conditions were monitored to detect the first event of failure (tensile fracture or strut buckling) around the crack front and flanks. The XCT images shown in Figs. 1, 2, 5 and S7 (and Supplementary Movies S1-4) were acquired using the long scan protocol with a 10 μm × 10 μm × 10 μm effective voxel size. To avoid XCT artefacts, beam energies were maintained below 10 W and only geometrical magnification was employed.

***Finite element (FE) calculations:*** Three-dimensional FE calculations were performed in the commercial package ABAQUS, using linear elastic material properties and in the small strain setting. The geometry of the specimens with embedded and through-thickness cracks used in the calculations was identical to the model used to 3D print the test specimens. Each strut in the octet-truss micro-structure was represented by Timoshenko beam elements (B31 in the ABAQUS notation) with a circular cross-section of radius $r_0$ and length $b = \ell/\sqrt{2}$ : the radius $r_0$ is related to the relative density via $\bar{\rho} \equiv 6\sqrt{2}\pi(r_0/b)^2$. Each strut was discretised via 24 B31 elements in a region of leading dimension $2a$ around the crack (to accurately capture the crack tip fields) while further away from the crack each strut was discretised by only 3 elements (mesh convergence studies confirmed the accuracy of the discretisation). This implied that the FE calculations had $20 - 100$ billion degrees of freedom. Uniaxial and multi-axial loading (at fixed $\lambda \equiv Q/P$) was applied by imposing monotonically increasing forces on the FE nodes on the specimen surfaces with all nodes on a given surface subjected to equal forces. Two types of analyses were conducted: (i) To investigate failure by strut tensile fracture, the specimens were loaded to failure load $P = P_T$ until a strut (typically in the close vicinity of the crack) attained the critical axial tensile stress $\sigma_f$; (ii) For the buckling failure mode, the eigenvalues were calculated for the imposed loading and the failure load $P_B$ set equal to the lowest eigenvalue. The operative failure mode is the one associated with the lowest failure load and hence $P_f = \min(P_T, P_B)$ with $K_{IC}$ then calculated by setting $P = P_f$ in (1). Details of the asymptotic calculations (Fig. 3a to 3d) are provided in Section B of the SI.

**Acknowledgements:** We acknowledge funding from the Office of Naval Research (N00014-18-1-2658 and N00014-20-1-2504:P00001). AS is supported by Cambridge-India Ramanujan scholarship from the Cambridge Trust and the SERB (Govt. of India). **Competing interests:** Authors declare no competing interests.; and **Data and materials availability:** All data is available in the main text or the supplementary materials.

**Additional information**

**Supplementary Information** is available for this paper.

# Supplementary Information

*Supplementary Text*

*A. Fabrication and characterisation of the octet-truss specimens*
*B. Mechanical properties of the parent material and octet-truss specimens*
*C. Description of the multi-axial loading setup and the measurement protocol*
*D. Calibration of $K_I$ and $T$ via FE calculations*
*E. Finite element calculations of the asymptotic fields at the crack-tip of the octet-truss*
*F. Fracture mechanism maps and calculations for other metamaterial topologies*
*G. Scaling laws and normalisation for the fracture mechanism map*

*Figs. S1-S13*
*Table S1 and S2*
*Captions for Movies S1 to S4*

*A. Fabrication and characterisation of the octet-truss specimens*

The octet-truss specimens used in the present study were fabricated using a novel large area stereolithography (SLA) technique (Methods). A 3D model was first built using a custom code and then sliced into 2D patterns. These 2D patterns are sequentially transmitted to a special light modulator, which is illuminated with UV light from a light emitting diode (LED) array. Each image is projected through a reduction lens onto the surface of the photosensitive resin. The exposed liquid cures, forming a layer in the shape of the 2D image, and the substrate on which it rests is lowered in steps of 5 μm so as to reflow a thin film of liquid over the cured layer. The image projection is then repeated, with the next image slice forming the subsequent layer (Fig. S1).

To enhance the capability of the fabrication setup to manufacture the architected metamaterials with ~10 million unit cells, the projector was coordinated with an optical scanning system. This allowed the production of large-scale parts (100 mm in overall size) but with micron scale resolution. As the mirrors move along the $x$-axis or $y$-axis, 2D patterns are reflected onto a new area next to the previously exposed area. The pattern change on the projector is coordinated with the scanning rate of the scanning mirror system. A customized focusing lens is used below the scanning optics to project the image onto the liquid surface. The frame is updated as the image is moved via the scanning optics to effectively create a continuous image in the photosensitive material. This technique allows the fabrication of polymer template architected material, hundreds of millimetres in size, with multiscale 3D architected features down to the 10 μm scale.

It is instructive to characterise the geometry of the as-printed octet-truss specimens to have an understanding of printing imperfections associated with layering artefacts, build orientation, heat dissipation and non-uniform crosslinking etc. associated with our sub-section projection SLA technique. We shall address the influence of imperfections in two steps: in this section we report a geometrical characterisation of the printed specimens to quantify visually observable imperfections and then in the following section we summarise the basic mechanical properties and relate them to the properties of the parent TMPTA in order to understand the role of imperfections on properties of interest.

Geometrical features of the $\bar{\rho} = 0.1$ octet-truss specimen with nominal strut length $b = 530$ μm and strut diameter $2r_0 = 65$ μm were investigated in detail via high resolution X-ray CT (effective pixel size $\sim 5$ μm). These cubic specimens of size $2W = 30$ mm comprised ~64,000 unit cells, ~1.5 million struts and ~0.5 million nodes and the aim of our analysis was to characterise imperfections and look for distortion of the polymerised 3D geometry *(1)* at the scale of nodes/struts For this purpose, we developed an in-house code that coupled to the commercial software VGStudioMax 3.4 to calculate strut diameters/node sizes by best fitting spheres at intervals of ~5 μm within the solid material of the octet-truss specimen. Image of the CAD drawing of the specimen used in the printing is shown in Fig. S2a along with the unit cell, an optical image of the as-printed specimen in Fig. S2b and the X-ray CT image of the specimen in Fig. S2c with the inset showing a single unit cell and a more detailed larger image in Fig. S2d. The X-ray CT images are shaded by the diameter $D$ of the best-fit sphere fitted at every location in the solid material. The data over the $\approx 10^7$ spheres fitted in the specimen is plotted as a distribution in Fig. S2e. The distribution is bimodal with the first mode corresponding to spheres fitted within struts and the second mode resulting from spheres fitted in nodal regions. The distributions corresponding to the two modes are clearly delineated with the mode of the distributions corresponding to the struts occurring at $D \approx 2r_0 = 65$ μm with a standard deviation $< 0.02D$. The second distribution corresponding to the nodal regions has a mode at $D \approx 76$ μm consistent with the node size in the CAD drawing.

While the above data demonstrates the high print quality with imperfections less than ~7 % over the whole specimen it is instructive to determine whether the printing process resulted in gradient over the specimen. For this purpose, we cut the above $2W = 30$ mm (Fig. S2f) as-printed specimen into 27 cubes each of side ~10 mm as shown in Fig. S2g. Each of these cubes was then imaged via high-resolution X-ray CT scans and the above procedure repeated (Figs. S2h and S2i). The measured probability distribution of the best-fit sphere diameters $D$ is included in Fig. S2i for all the 27 sub-cubes with the shaded region showing the variation over the 27 sub-cubes and solid line the mean over all the sub-cubes. Clearly there is very little geometrical variability of the specimen as a result of printing induced imperfections *(1,2)*. This procedure was repeated for specimens over the range of $\bar{\rho}$ values investigated here with the results showing very similar trends as those seen in Fig. S2i.

*A1.    Embedded crack geometry in the octet-truss specimens*
Plan view (i.e., view perpendicular to the plane of the cracks) sketches of the circular and square crack geometries in the octet-truss specimens are included in Figs. S6a and S6b, respectively for cracks of size $a/\ell = 4$, where $2a$ is either the diameter of the circular crack or the side of the square crack and $\ell$ the cell size. The cubic geometry of the octet-truss unit cell implies that a "perfect" square crack geometry can be constructed. However, the cubic unit cell implies that a circular crack shape can only be approximated via a jagged crack front as shown in Fig. S6c. The deviation from a perfect circular crack front becomes increasingly severe with decreasing $a/\ell$ with circular crack fronts shown in Fig. S6d for three values of $a/\ell$. However, it is clear that to a reasonable approximation a circular crack is still clearly definable for $a/\ell = 4$ which is the smallest value of $a/\ell$ considered in this study: below $a/\ell = 4$ the notion of a crack becomes tenuous, and it is more appropriate to think of the defect as a set of missing struts.

We note in passing that while the octet-truss geometry not only lends itself more readily to define a square crack as seen in Fig. S6 but also, we show in Section D that such as square crack results in a near constant value of the stress intensity factor $K_I$ along the crack front (Fig.

S9b) allowing for easier interpretation of toughness measurements. These two reasons were the key driver behind our choice of an embedded square crack in this study.

### B. *Mechanical properties of the parent material and octet-truss specimens*

Here we report mechanical property measurements including those of the bulk TMPTA, single printed struts and the octet truss specimens to investigate the role of printing induced imperfections and confirm that the properties of the printed specimens are consistent with well-established relations between the octet-truss and parent material properties.

### B1. *Mechanical characterisation of the parent trimethylolpropane triacrylate (TMPTA)*

Dog-bone specimens were fabricated in different print directions made from trimethylolpropane triacrylate (TMPTA) used as the parent material for the manufacture of the octet-truss metamaterials (Fig. S4a). The tensile response was measured over three orders of magnitude in applied strain rate ($10^{-3} s^{-1} \leq \dot{\varepsilon} \leq 10^{-1} s^{-1}$) with the stress defined via the load measured from the load cell of the test machine and strain defined by measuring displacement using a laser extensometer over the central gauge-length of the dog-bone specimen. Three repeat tests were performed in each print direction and results plotted in Fig. S4b with the shaded zone showing the variation over the 9 tests that were performed and the dark line the average of these measurements.

It is also instructive to confirm that these bulk parent material properties are valid at the scale of the struts within the octet truss specimen. For this purpose, we printed a sandwich specimen as shown (CAD model shown in Fig. S4c while optical images are included in Fig. S4d and e) comprising a $10 \times 10$ cubic array of vertical struts of diameter 45 µm, height $h_0 = 540$ µm and spaced ~700 µm apart. The specimen including the ~250 mm thick face-sheets were printed along in a single print with the print direction aligned with the height of the struts as shown in Fig. S4e. The face-sheets were then glued to the platens of the tensile test machine and the tensile response measured at an applied strain rate $\dot{\varepsilon} = 10^{-3}$ s$^{-1}$. The load $P$ was measured using a 10 N load cell while strain in the strut defined by measuring the imposed displacement $u$ between the face-sheets of the specimen using a laser extensometer. The measured stress $\sigma \equiv P/(NA_0)$ where $N = 100$ is the number of struts in the specimen and $A_0$ the cross-sectional area of a single strut versus strain $\varepsilon \equiv u/h_0$ is included in Fig. S4b and is remarkably similar to that of the bulk printed TMPTA indicating that printing of micron-sized struts does not affect the TMPTA properties in any significant manner. Both these bulk TMPTA and strut tests consistently show that the TMPTA behaves as a linear elastic material with a Young's modulus $E_s = 430$ MPa and tensile strength $\sigma_f \approx 11$ MPa. This linear elastic brittle fracture behaviour is consistent for the entire range of strut dimensions (Table S1) and do not display any size effects on mechanical properties *(3-5)* unlike that reported for micro-pillars *(3)* and nanolattices *(4)*. Another important mechanical property is the fracture toughness of the parent TMPTA. We printed rectangular specimens of height $H = 30$ mm, width $W = 10$ mm and thickness $B = 3$ mm with edge cracks of sizes in the range 0.2 mm ≤ $a$ ≤ 3 mm. This sharp crack was created by first using a wire cutter of 0.3 mm blade thickness to create a notch and then tapping in a fresh razor blade following ASTM D5045: the crack size $a$ is defined to the tip of the sharp crack created by the razor blade and measured via an optical microscope. The specimens were gripped using wedge grips and subjected to tensile loading at an imposed displacement rate 0.5 mm min$^{-1}$ along the specimen height $H$ as shown in Fig. S12a. The applied load $P$ was measured via the load cell of the test machine and the load $P = P_f$ at fracture recorded. Measurements of the failure stress $\sigma_{\text{frac}} \equiv P_f/(WB)$ as a function of the crack size $a$ are included in Fig. S12b on a log-log plot: $\sigma_{\text{frac}} \propto 1/\sqrt{a}$ for $a = a_T \geq 0.9$ mm with $\sigma_{\text{frac}} \approx \sigma_f \approx$ 11 MPa for $a < a_T$. This clearly suggests that failure of the edge-cracked specimen is toughness governed above the transition flaw size of $a_T \approx 1$ mm and material strength limited for smaller flaws. From the data $a > a_T$ we can calculate the fracture toughness of the TMPTA via

$$K_{IC} = Y_I(a/W)\sigma_{\text{frac}}\sqrt{\pi a}, \tag{B1}$$

where

$$Y_I = 1.12 - 0.23\frac{a}{W} + 10.6\left(\frac{a}{W}\right)^2 - 21.7\left(\frac{a}{W}\right)^3 + 30.4\left(\frac{a}{W}\right)^4. \tag{B2}$$

The measurements give $K_{IC} = 0.6$ MPa$\sqrt{\text{m}}$.

The transition flaw size of $a_T = 1$ mm is of significant importance in terms of the design of our octet-truss specimens. Table S1 gives the range of the maximum/minimum relative densities, strut lengths and diameters of the specimens employed in this study. The maximum strut diameter of 65 μm $\ll a_T$ so that it is clear that tensile fracture of crack tip struts in the octet-truss specimens is purely governed by the material strength $\sigma_f$ and surrounding struts remain purely elastic with no damage. Thus, toughness of the octet-truss specimen is set via purely elastic fracture with no associated process (damage) zone or even small-scale yielding.

*B2. Effective properties of the printed octet truss specimens*
To check the consistency of the effective properties of the printed octet-truss specimens we performed tensile tests on specimens without cracks in three orthogonal directions. These tests were performed in a manner identical to the fracture tests except that the cubic specimen of size $2W$ did not have an embedded crack (Fig. S3a). The cubic symmetry of the octet-truss implies that in the absence of printing induced anisotropies the tensile responses should be identical along the $x, y$ and $z$ −directions (Fig. S3b) which are aligned with the [110], [$\bar{1}$10] and [001] directions, respectively of the fcc microstructure. Stress $\sigma_{aa}$ is defined as the ratio of the load to cross-section area while strain $\varepsilon_{aa}$ is the ratio of the imposed displacement to the specimen size $2W$ with $a$ taking values $x, y, z$ depending on the direction of loading. Measurements of $\sigma_{aa}$ versus $\varepsilon_{aa}$ for the $\bar{\rho} = 0.1$ and 0.02 specimens are included in Figs. S3c and S3d and show that to within 5 % the responses are identical in the three orthogonal directions demonstrating that there are no significant printing induced anisotropies *(6)*. Also included in Figs. S3c and S3d are predictions of the tensile responses based on the measured bulk properties of the TMPTA based on predictions in *(7)*; viz. the Young's modulus of the octet-truss

$$E_{xx} = E_{yy} = E_{zz} = \frac{1}{9}\bar{\rho}E_s \tag{B3}$$

and the failure stresses

$$\Sigma_{\text{failure}} = \begin{cases} \frac{1}{3}\bar{\rho}\sigma_f & \bar{\rho} \geq \frac{24\sqrt{2}}{n^2\pi}\left(\frac{\sigma_f}{E_s}\right) \\ \frac{n^2\pi\bar{\rho}^2}{72\sqrt{2}}E_s & \text{otherwise,} \end{cases} \tag{B4}$$

where $n \approx 1.9$ defines the buckling mode of the low relative density octet-truss specimens and calculated via a FE eigenvalue analysis. These predictions are included in Fig. S3d and not only correctly predict that the $\bar{\rho} = 0.1$ and 0.02 specimens fail by tensile strut fracture and elastic strut buckling, respectively but also are in excellent quantitative agreement. This confirms that the strut mechanical properties of the printed octet-truss specimens are consistent with bulk TMPTA and that in line with the direct XCT observations reported in Section A, the imperfections in the printed specimens are sufficiently small so as to have a negligible effect on the mechanical properties.

*B3. Mechanical variability within an octet-truss specimen*
To examine for variability in mechanical properties over the printed octet-truss specimens comprising ~64,000 unit cells we cut the $2W = 30$ mm specimens into 27 sub-cubes as discussed in Section A in the context of the X-ray imaging (Figs. S3e and S3f). These specimens were then tested in uniaxial tension (loaded in the print direction) as shown in Fig. S3g using the procedure described above for the full specimens. The measured tensile stress $\sigma_{zz}$ versus $\varepsilon_{zz}$ curves are included in Figs. S3h and S3i for $\bar{\rho} = 0.1$ specimens with two different choices of unit cell dimensions. The shaded regions show the variability over the 27 separate cubes and the solid lines the corresponding means. In line with the minimal geometrical variations over the specimen (Fig. S2i), the mechanical properties also show minimal variability over the specimen indicating that the printing process does not induce gradients in mechanical properties within a specimen.

## C.  Description of the multi-axial loading setup and the measurement protocol

*C1. Experimental rig*
The loading system (Fig. 1d and 2a) used to measure fracture of the 3D octet-truss specimens with embedded cracks was designed such that XCT scans could be acquired while the specimen was subjected to multiaxial loads with varying triaxialities $\lambda$. We designed an in-situ XCT rig (Fig. S8) for the 30 mm × 30 mm × 30 mm specimen with peak load capacity of 5 kN in each of the three orthogonal directions. The rig was built using Perspex blocks and Nylon screws to avoid high energy scans and subsequent beam hardening corrections. The rig when placed inside the scanner had base dimensions 200 mm × 200 mm and was 300 mm high. Each loading axis consisted of a load cell, a linear spring to ensure contact was maintained with the specimen and a linear stepper motor for the application of the force. To apply the loads, the three stepper motors were interfaced to an Arduino Uno Rev3 board. Data from the load cells were logged on the same control board with a reading frequency ten times excitation frequency of the stepper motors. The combined data logger and stepper motor control system interfaced with a control algorithm which fed back into the Arduino board to ensure that the load ratio was maintained at the prespecified level. A linear voltage displacement transducer (LVDT) was used to measure displacements in the $z-$direction. Specimens were prepared for the triaxial tests by gluing to platens having wedged edges as shown in the inset of Fig. S8. For uniaxial loading, the in-plane loading axes of the setup were dismantled.

*C2. Uniaxial and triaxial loading protocol*
The fracture toughness measurements involved both uniaxial and multiaxial loadings. During uniaxial testing, displacement-controlled loading was applied via steps of linear displacement through the stepper motor. To ensure contact, smooth loading and avoid jerks on the specimen a range of spring stiffness were used from 0.34 N/mm to 50 N/mm. They were chosen based on the stiffness of the specimen which is mainly governed by the relative densities. The load-line displacement shown in Fig. 1e was measured from one such an experiment on a $\bar{\rho} = 0.08$ specimen. For the uncracked metamaterial, loads were applied in steps of 0.05 mm and the actual displacements on the specimen read off using the LVDT. Load $P$ was continuously read against the applied displacement $u$ as plotted in Fig. 1e. For the cracked specimen, XCT scans were taken at regular intervals of loading, as described in C3. The loading was continued until the specimen fails or when the load drops due to an observed crack failure mode (Fig. 1, 2 and S8). The triaxial loading involved two additional axes and the requirement of maintaining a constant triaxiality $\lambda$ throughout the entire test. This was achieved by controlling the displacements of stepper motors based on the loads read-off from the z-axis, and the planar axes load cells marked as $P$ and $Q$, respectively in Fig. 2a. This information was fed into the Arduino board to control the displacements and hence maintain a constant triaxiality $\lambda$. The reading frequency from the load cell was adjusted to be ten times the excitation frequency of the stepper motors for precise adjustments in loading. This feedback and control mechanism ensured that the load ratio was maintained at the prespecified levels over the range of different triaxialities $\lambda$ shown in Fig. 2b.

*C3. Protocols for XCT scans*

The experimental rig was integrated with the XCT scanner for in-situ observation during the loading (Fig. S8). There were two major challenges. First, the physical dimension of the scan area (with the experimental rig) increased nearly tenfold as compared to the standalone specimen. This resulted in loss of spatial resolution. Second, the in-plane loading axes interrupted the line of sight of the X-ray source and significantly attenuated the beam intensities. A typical scan therefore required high beam energies, long scan times and resulted in large effective voxel sizes. To overcome these limitations two different scan protocols were followed for each test. The scanner was first calibrated to acquire 2D projections for a small opening window (between $50°$ of specimen rotation) viewing only the crack and avoiding the loading axes. Optical magnification was added to compensate for the loss in spatial resolution. Using the above configuration and optimal scan parameters (Table S2), the acquisition times were reduced to 8-15 mins. A quick scan protocol was followed to continuously monitor the crack conditions at regular intervals during loading. At the detection of the first failure event, the in-plane loadings were arrested, and the peripheral assembly dismounted. The rig was then moved closer to the X-ray source for higher magnification without the need for optical magnification. The specimen was rotated a full $360°$ for enhanced resolution and artefact free scans. This scan protocol was called the long scan and typically took about an hour (Table S2). Prior to the tests, each specimen was investigated via the long scan protocol to check fabrication quality as parameterised by strut imperfections and deviation of the as-printed dimensions compared to the CAD input model. Uniformity in quality of the specimens was ensured before they were tested. In-house MATLAB codes were used to reconstruct the 3D volumes in the quick scans, while all other measurements like relative densities and dimensioning were performed in the commercial package VGStudioMax 3.4. Images and videos are presented as acquired from the scans with no further post-processing in graphical or rendering software.

**D.     *Calibration of $K_I$ and $T$ via FE calculations***

The octet-truss lattice has cubic symmetry and for low relative densities (i.e. $\bar{\rho} \leq 0.2$), the stiffness of the lattice is dominated by the axial stiffness of the constituent struts. In this limit, elastic stress versus strain relation in the $(x, y, z)$ co-ordinate system (Fig. 1a) is given by *(7)*

$$\begin{pmatrix} \varepsilon_{xx} \\ \varepsilon_{yy} \\ \varepsilon_{zz} \\ \gamma_{yz} \\ \gamma_{xz} \\ \gamma_{xy} \end{pmatrix} = \frac{1}{\bar{\rho} E_s} \begin{pmatrix} 9 & -3 & -3 & 0 & 0 & 0 \\ & 9 & -3 & 0 & 0 & 0 \\ & & 9 & 0 & 0 & 0 \\ & & & 12 & 0 & 0 \\ & \text{sym} & & & 12 & 0 \\ & & & & & 12 \end{pmatrix} \begin{pmatrix} \sigma_{xx} \\ \sigma_{yy} \\ \sigma_{zz} \\ \sigma_{yz} \\ \sigma_{xz} \\ \sigma_{xy} \end{pmatrix}, \qquad (D1)$$

in terms of the parent material Young's modulus $E_s$ with the engineering shear strains $\gamma$ related to strains $\varepsilon$ via $(\gamma_{yz}/2, \gamma_{xz}/2, \gamma_{xy}/2) \equiv (\varepsilon_{yz}, \varepsilon_{xz}, \varepsilon_{xy})$. These analytical relations were calculated assuming pin-jointed struts which is expected to be at low relative densities. To check their accuracy for the 3D lattice printed here we performed unit FE calculations using the CAD drawing used in the printing process (Fig. S13 a)). The geometry was discretized using C3D20R elements of size $\sim 0.25 r_0$ and predictions of the Young's modulus $E_{zz}$ and the Poisson's ratio $\nu_{zx}$ are included in Figs. S13b and S13c, respectively as a function of the relative density $\bar{\rho}$. The corresponding analytical predictions from (D1) are also included and show the excellent accuracy of the pint-jointed strut model for $\bar{\rho} < 0.2$ *(8)* which is within the whole range of relative densities investigated here. We therefore use the analytical elasticity relation (D1) in all our analyses.

Calibration of the $K_I$ and $T$ terms in the asymptotic expansion (2) was carried out via FE calculations using the commercial package ABAQUS 6.14 where the cracked octet-truss specimen was modelled as smeared out elastic continuum with the anisotropic elastic relation (D1). To analyse the experiments two geometries were considered: (i) the 3D embedded crack test specimen (Fig. 1a) and (ii) the centre-cracked specimen with a through-thickness crack (Fig. 5a). In addition, a 3D 4-point bend specimen was also analysed for the case study reported in Fig. 4 along with situations involving complex defects geometries (Fig. S10). Recall that in the continuum elastic representation there is no microstructural length scale geometrical non-dimensional length scales (such as $a/W$ for the specimen in Fig. 1a) fully describes the cracked geometry. The 3D specimens were discretised using 8-noded brick elements with reduced integration (C3D8R in the Abaqus notation) while a 2D plane-strain representation was employed for the specimens with through-thickness cracks. These 2D specimens were discretised using 4-noded plane-strain elements with reduced integration (CPE4R in the ABAQUS notation). Typically, elements of size $a/200$ were used in a region of size $2a$ around the crack tip/front to accurately capture the crack tip stress-fields.

*D1. 3D embedded crack test specimens*
A mathematically sharp square crack with sides aligned with the global $(x, y, z)$ co-ordinate system (Fig. S9a) was analysed. The cubic specimens of side $2W$ also had its edges aligned $(x, y, z)$ (Fig. S9a) so that stresses and strains in the global co-ordinate system of the cubic specimen are also related via (D1). First consider the case of uniaxial loading with a tensile load $P$ in the $z-$direction (perpendicular to the crack plane) and $Q = 0$ (Fig. 2a). To calculate the resulting $K_I$ along the crack front we evaluated the J-integral *(9)* using the virtual crack extension/domain integral method *(10,11)*. Consider the crack front co-ordinate system $e_i$ where $e_1 - e_3$ forms the crack plane and $e_1$ is normal to the crack front as shown in the inset of Fig. S9a. We restrict attention to the case where crack growth occurs within the crack plane and normal to the crack front and calculate the J-integral as a function of the position $\zeta$ along the crack front, where $\zeta = 0$ is located at the mid-point along the side of the crack (inset of Fig. S9a). The J-integral $J$ is then related to the mode-I, mode-II and mode-III stress intensity factors $K_I$, $K_{II}$ and $K_{III}$, respectively via

$$J = \frac{1}{8\pi}[K_I B_{11}^{-1} K_I + K_{II} B_{12}^{-1} K_{II} + K_{III} B_{13}^{-1} K_{III}], \tag{D2}$$

where $B_{ij}$ is the pre-logarithmic energy factor matrix for a single dislocation parallel to the crack front *(12)*. In order to calculate the J-integral accurately we employed an FE mesh with elements of size $a/200$ in a region of size $2a$ around the crack tip/front, where $2a$ is the crack size. While the J-integral is theoretically path independent it is well known that the contours with radii $r < 10^{-2}a$ do not provide accurate results because of numerical singularities. Therefore, we only considered contours with $r > 10^{-2}a$ which all gave the same value of the J-integral to within 0.05%.

In general, $K_I$, $K_{II}$ and $K_{III}$ are calculated by the interaction integral method *(13)*. For our case of tensile loading perpendicular to the crack plane, symmetry requires $K_{II} = 0$ with $K_{III}$ also vanishing at $\zeta = 0$. We define the mode-I and mode-III stress-intensity calibration factors via the relations

$$K_I = Y_I(\zeta/a) \frac{P}{4W^2} \sqrt{\pi a}, \tag{D3}$$

and

$$K_{III} = Y_{III}(\zeta/a) \frac{P}{4W^2} \sqrt{\pi a}, \tag{D4}$$

respectively. Predictions of $Y_I$ and $Y_{III}$ as a function of the normalised crack-front position $\zeta/a$ are included in Fig. S9b. The mode-I calibration factor $Y_I$ is approximately independent of $\zeta$ and for $\zeta \approx 0$, $Y_{III} \ll Y_I$. In the study we restrict attention to the initiation of crack growth at $\zeta = 0$ where $Y_I \approx 2.2/\pi$ and $Y_{III} = 0$. In the limit at $\zeta = 0$, and the crack orientation as shown in Fig. S6a, $J$ and $K_I$ can be related by the simple relation

$$J = \frac{K_I^2 \sqrt{6}}{E_s \bar{\rho}}, \tag{D5}$$

by assuming the strain in the crack front co-ordinate system $\varepsilon_{33} = 0$. The full interaction integral calculations revealed that the simple relation (D5) is accurate to within a few percent thus illustrating that the approximation $\varepsilon_{33} = 0$ at the crack front holds to a high level of accuracy. Thus, (D5) serves as a useful analytical approximation.

The addition of loads $Q$ has no effect on $K_I$, $K_{II}$ and $K_{III}$ and thus the calibration factors in Fig. S9b are independent of $\lambda$. In Fig. S9b we also included predictions of $Y_I$ assuming an isotropic elastic medium with Poisson's ratio 0.3: as is well known *(14)* the dependence of $Y_I$ on position is significantly stronger in the isotropic solid compared to the anisotropic cubic material representative of the octet-truss metamaterial. In fact, the choice of the square crack geometry for the octet-truss specimens was driven by this fact that the elastic anisotropy of the octet-truss results in $K_I$ being approximately constant along the square crack edge which enables easier interpretation of the measurements.

Now consider the calibration of the so-called T-stress under uniaxial load $P$. Before discussing the calibration, it is first worth revisiting the asymptotic expansion of the stress-field near a mathematically sharp crack. In the crack front co-ordinate system $e_i$ (inset of Fig. S9a) the stress $\sigma_{ij}$ at a location given by the polar co-ordinates $(r, \theta)$ in a plane normal to the crack front (Fig. 3a) is

$$\begin{aligned}\sigma_{ij} = r^{-1/2}\big(K_I f_{ij}^I + K_{II} f_{ij}^{II} + K_{III} f_{ij}^{III}\big) + T_{11}\delta_{1i}\delta_{1j} + T_{33}\delta_{3i}\delta_{3j} + T_{13}\delta_{1i}\delta_{3j} \\ + \mathcal{O}(r^{\frac{1}{2}}) + \mathcal{O}(r) + \dots,\end{aligned} \tag{D6}$$

where $\delta_{ij}$ is the Kronecker delta, $(T_{11}, T_{33}, T_{13})$ the T-stresses and $(f_{ij}^I, f_{ij}^{II}, f_{ij}^{III})$ are non-dimensional functions that give the angular dependence of the stresses around the crack and depend on the elastic constants and $\theta$. The expansion shows that the T-stresses are defined as terms in the expansion that give a spatially constant contribution to $\sigma_{ij}$. Normalizing (D6) using definitions $\overline{K}_I \equiv K_I/(\sigma_f \sqrt{\ell})$, $\overline{T}_{ij} \equiv T_{ij}\sqrt{\ell}/K_I$ and $\bar{r} \equiv r/\ell$ we recover (2) where we have neglected the contributions from the $K_{II}$ and $K_{III}$ terms. We shall restrict ourselves to calculating the T-stresses for crack growth normal to the crack front at $\zeta = 0$. Then, under uniaxial loading (Fig. S9a), symmetry requires $T_{13} = 0$. Moreover, to a very high level of accuracy the FE calculations revealed that the strain $\varepsilon_{33} = 0$ at $\zeta = 0$ as discussed above. This reduces the problem to calibrating $T = T_{11}$. The T-stress along three-dimensional crack fronts in isotropic elastic solids can be evaluated using an interaction integral *(15)*. To the best of the authors' knowledge, there exists no equivalent method for evaluating the T-stress along three-dimensional crack fronts in anisotropic elastic solids. We thus used the following approximate method. Recall that at $\zeta = 0$, $K_{II} = K_{III} = 0$ and thus in the crack-front co-ordinate system $e_i$ we write the stress $\sigma_{11}$ along $\theta = 0$ as

$$\sigma_{11} = \frac{c}{\sqrt{r}} + T_0 + \mathcal{O}(r^{\frac{1}{2}}) + \mathcal{O}(r), + \ldots, \tag{D7}$$

where $c$ is a constant related to $K_I$ and $T_0$ denotes the T-stress for uniaxial loading with $Q = 0$. For $r \to 0$ it suffices to neglect the terms of order $\mathcal{O}(r^{\frac{1}{2}})$ and higher. We then used the FE calculations to determine $\sigma_{11}$ over the range $10^{-2} \leq r/a \leq 10^{-1}$ and used a regression analysis to fit the function $\sigma_{11} = c/\sqrt{r} + T_0$ to determine $c$ and $T$ for a given applied $P$. The outcome of this analysis was $\hat{T}_0 \equiv T_0\sqrt{a}/K_I \approx -\sqrt{\pi}/2.2$. Now consider the case of multiaxial loading with a triaxiality $\lambda \equiv Q/P$. The total T-stress is then given by linear superposition as $T = T_0 + Q/(4W^2)$. Substituting for $K_I$ from (D3) and noting that $Y_I \approx 2.2/\pi$ we obtain

$$\hat{T} \equiv \frac{T\sqrt{a}}{K_I} \approx \frac{\sqrt{\pi}}{2.2}(\lambda - 1). \tag{D8}$$

Then (3) follows from the definition $\overline{T} \equiv \hat{T}\sqrt{\ell/a}$.

### D2. Specimens with through-thickness cracks

We analysed a centre-cracked panel in plane-strain ($\varepsilon_{yy} = 0$) with $z-$direction normal to the crack plane as shown in Fig. 5a. The specimen was loaded in uniaxial tension in the $z-$direction and the mode-I stress-intensity factor evaluated for straight-ahead crack-growth. The stress-intensity factor was again determined by evaluating the J-integral and related to $K_I$ via (D2). In terms of the remote tensile stress $\sigma_\infty$, we can write $K_I$ as

$$K_I = Y_I \sigma_\infty \sqrt{\pi a}, \tag{D9}$$

and the FE calibration gave $Y_I \approx 1$ for the $a/W = 0.2$ specimens used in this study.

### D3. The 4-point bend case of Fig. 4

In Fig. 4 we present a case study illustrating the fracture analysis of a cracked structure comprising the octet-truss metamaterial. For this purpose, we consider a 3D beam of cross-section $2W \times 2W$ under 4-point bending with the longitudinal axis of the beam aligned with the global $y-$direction and the global $z-$direction aligned with the transverse axis (Fig. 4a). A square crack of side $2a$ is present at mid-section with the plane of the crack perpendicular to the longitudinal axis of the beam. The crack is symmetric about the neutral axis of the beam as shown in Fig. 4a while the cubic directions of the octet-truss microstructure are aligned with the global $(x, y, z)$ directions. The aim of the case study is to illustrate the calculation of the failure load $P_f$ without performing FE simulations of the detailed octet-truss microstructure. Rather we perform FE calculations using a smeared-out continuum (Fig. S11a) (which require significantly fewer degrees of freedom compared to calculations with the detailed microstructure) to evaluate appropriate loading parameters which can then be combined with the fracture mechanism map (Fig. 3e) to estimate $P_f$. The idea here is that the fracture mechanism map includes all detailed microstructural information of fracture and those expensive calculations need not be repeated with the map serving as a lookup chart of the fracture properties of the octet-truss metamaterial.

The smeared-out continuum used to represent the octet-truss metamaterial is the anisotropic elastic relation (D1). For the beam under 4-point bending, the moment in the central-section of length $L = 20W$ is spatially uniform and given by $M = Ps$. Thus, it suffices to analyse a beam of length $L$ and square cross-section subjected to a pure moment $M$ as shown in Fig. S11b. The origin of the co-ordinate system $(x', y', z')$ of this section is located at the beam centre as shown in Fig. S11b and the moment was applied via an imposed displacement field $u_{y'} = \mp \kappa z'$ on the ends $y' = \pm L/2$, where $u_{y'}$ is the imposed displacement in the $y'$−direction corresponding to a beam curvature $\kappa$. The applied moment is then evaluated by integrating the tractions over the area $A = 4W^2$ of the beam cross-section at $y' = L/2$, i.e.

$$M = \int_A T_{y'} \, z' dz' dx', \tag{D10}$$

where $T_{y'}$ is the traction in the $y'$−direction.

The inset of Fig. S11b shows the crack front along with the local crack front co-ordinate system $e_i$. In line with the embedded crack study (Section D1) we restrict attention to the case where crack growth occurs in the crack plane ($e_1 - e_3$ plane) and normal to the crack front. Again, symmetry requires that $K_{II} = 0$ and $K_I$ and $K_{III}$ were evaluated by first calculating the J-integral and then using the interaction integral method to estimate $K_I$ and $K_{III}$ as described in Section D1. Given the moment loading $M = Ps$, we write the stress-intensity factors as

$$K_I = Y_I(\zeta/a) \frac{Ps}{8W^3} \sqrt{\pi a}, \tag{D11}$$

and

$$K_{III} = Y_{III}(\zeta/a) \frac{Ps}{8W^3} \sqrt{\pi a}. \tag{D12}$$

The FE predictions of $Y_I$ and $Y_{III}$ as a function of normalised location $\zeta/a$ along the crack front are shown in Fig. S11c. Similar to the cubic specimens of Section A1, $Y_I$ is relatively insensitive to $\zeta/a$ and $Y_{III} \ll Y_I$. It is thus sufficient to focus on mode-I crack growth along $\zeta = 0$. Given that we are restricting ourselves to analyzing crack growth normal to the crack front at $\zeta = 0$, a symmetry argument similar to that of Section A1 implies that the T-stress problem reduces to calibrating $T = T_{11}$. This analysis was done in the manner described in Section A1, i.e. we fitted the function $c/\sqrt{r} + T$ to the FE predictions of the variation of $\sigma_{11}$ with distance $r$ along the $e_1$−direction. The calibration function $\hat{T}$ is then defined as $\hat{T} \equiv T\sqrt{a}/K_I$ and $\hat{T}$ is predicted to be 1.13.

With these calibrated values of $Y_I$ and $\hat{T}$ the failure load $P_f$ can be predicted by using the fracture mechanism map (Fig. 3e) as detailed in the manuscript without having to resort to calculations with the full microstructural details of the octet-truss metamaterial. This brings an enormous reduction in the computational cost associated with performing a structural failure calculation. For example, for the simple structural problem of 4-point bending of a beam considered in Fig. 4a the smeared-out continuum FE calibration calculations required about a million degrees of freedom (DOFs). By contrast we estimate that modelling the detailed octet-truss microstructure to predict its failure would involve in excess of 200 billion DOFs. Of course, a full-scale structure made from a metamaterial might be 1000s of mm in size and will require trillions of DOFs and many terabytes of RAM to model all the architectural features.

*D4. The complex defect geometries of Fig. S10*

Calibration of $Y_I$ and $\hat{T}$ for any arbitrary loading/defect geometry can be performed in the manner described above and then used to estimate failure loads for structures made from metamaterials. In Fig. S10 we give some examples of such calibrations for structures made from the octet-truss metamaterial. The geometries analysed include:

(i) A penny-shaped crack of radius $a$ under multi-axial loading. $(Y_I, \hat{T})$ are defined as in Section A1 and are reasonably independent of the orientation $\phi$ (Fig. S10a) along which crack growth is assumed to occur. The predictions in Fig. S7a are valid for $a/W \leq 0.4$.

(ii) A circular crack emanating a distance $a$ from a spherical void of radius $r$ under multiaxial loading. Again $(Y_I, \hat{T})$ are defined as in Section A1 with predictions of these calibration factors given as a function of $a/r$ in Fig. S10b. Note that we have plotted a modified function of $\hat{T}$ whose value is independent of the load triaxiality $\lambda$. The predictions are valid for $(a + r)/W \leq 0.4$.

(iii) The beam under 4-point bending described in Section A3. Here however we allow for the crack to be offset a distance $\xi$ from the neutral axis as shown in Fig. S10c and included predictions of $(Y_I, \hat{T})$ as a function of the normalised offset $\xi/a$. The predictions in Fig. S7c are valid for $a/W \leq 0.2$.

(iv) The square crack within specimen under multi-axial loading. This is precisely the experimental specimen used in this study and detailed elsewhere. This case in included in Fig. S10d for the sake of completeness and also to emphasize that the calibrated values of $(Y_I, \hat{T})$ hold for $a/W \leq 0.4$ as indicated in Fig. S10d.

**E. Finite element calculations of the asymptotic fields at the crack-tip of the octet-truss**

To investigate the effect of T-stresses on the fracture of the octet-truss we performed an asymptotic analysis (boundary layer analysis) *(16)* on an octet-truss specimen that is one-unit cell thick (Fig. 3a). The plane $e_1 - e_2$ of deformation is parallel to the (010) plane of the fcc microstructure and $e_1$ parallel to the $[\bar{1}10]$ direction such that the planes $e_3 = 0$ and $e_3 = -\ell$ bound the section analysed. We imposed a displacement field comprising a combination of the mode-I stress intensity factor $K_I$ and the T-stress on the outer boundary to simulate the fields as you asymptotically approach the crack front within the specimens investigated here.

The section of the octet-truss specimen analysed comprised $50 \times 50$ unit cells in the $e_1 - e_2$ plane with the crack running along the $e_1$ −direction half-way through the square section and formed by removing a one layer of unit cells (Fig. 3a). This forms the sharpest possible crack within the octet-truss (Fig. S5). Convergence studies confirmed that the size of the region analysed is sufficiently large for the results to be independent of the domain size. For clarity we shall refer to the points of intersection of the struts of the octet-truss as vertices and all displacement boundary conditions discussed subsequently were only applied to these vertices with the displacements of FE nodes within the struts being outcomes of the solution. To simulate plane-strain we set the displacement $u_3 = 0$ for all vertices along with the displacement jumps $\Delta u_1 = \Delta u_2 = 0$ and rotation jumps $\Delta \phi_1 = \Delta \phi_2 = \Delta \phi_3 = 0$ (where $\phi_1, \phi_2, \phi_3$ are the rotations about the $e_1, e_2$ and $e_3$ axes, respectively) between corresponding vertices on the $e_3 = 0$ and $e_3 = -\ell$ planes. In order to simulate the effect of the $K_I$ −field and T-stresses in this asymptotic analysis define a polar co-ordinate system $(r, \theta)$ as shown in Fig. 3a centred at the mid-section of the crack-tip with $\theta = 0$ the direction of straight-ahead crack growth. On the outer boundary of the square-section analysed we then imposed displacements consistent with the asymptotic $K_I$ and T-stress fields in a solid with the cubic elastic relation (D1). These displacements are specified as follows *(17)*.

Recall that the $e_3$ −direction plane-strain direction is co-incident with the $x$ −axis of the specimen. We thus rewrite the compliance relation (D1) in the $e_i$ co-ordinate system, i.e.

$$\begin{pmatrix} \varepsilon_{11} \\ \varepsilon_{22} \\ \varepsilon_{33} \\ \gamma_{23} \\ \gamma_{13} \\ \gamma_{12} \end{pmatrix} = \frac{1}{\bar{\rho}E_s} \begin{pmatrix} 9 & -3 & -3 & 0 & 0 & 0 \\ & 9 & -3 & 0 & 0 & 0 \\ & & 9 & 0 & 0 & 0 \\ & & & 12 & 0 & 0 \\ & \text{sym} & & & 12 & 0 \\ & & & & & 12 \end{pmatrix} \begin{pmatrix} \sigma_{11} \\ \sigma_{22} \\ \sigma_{33} \\ \sigma_{23} \\ \sigma_{13} \\ \sigma_{12} \end{pmatrix}, \quad (E1)$$

where the cubic symmetry implies that the compliance matrix remains numerically unchanged. In more succinct Voigt notation (E1) reads $\varepsilon_i = a_{ij}\sigma_j$, where $(i,j) = 1,\ldots,6$ and $\varepsilon_i \equiv (\varepsilon_{11}, \varepsilon_{22}, \varepsilon_{33}, 2\varepsilon_{23}, 2\varepsilon_{13}, 2\varepsilon_{12})^T$ and $\sigma_i \equiv (\sigma_{11}, \sigma_{22}, \sigma_{33}, \sigma_{23}, \sigma_{13}, \sigma_{12})^T$. We then define plane-strain components of the compliance matrix as

$$b_{ij} = a_{ij} - \frac{a_{i3}a_{j3}}{a_{33}} \quad (i,j = 1,2,6), \quad (E2)$$

and an associated characteristic equation

$$b_{11}\mu^4 - 2b_{16}\mu^3 + (2b_{12} + b_{66})\mu^2 - 2b_{26}\mu + b_{22} = 0. \quad (E3)$$

Let $\mu_1$ and $\mu_2$ denote the complex roots of (E3) with $\text{Im}(\mu_1)$ and $\text{Im}(\mu_2) > 0$. Then in terms of $\mu_1$ and $\mu_2$ the displacement field applied on the boundary of the specimen is given by

$$u_1 = K_I \sqrt{\frac{2r}{\pi}} \text{Re}\left[\frac{1}{\mu_1 - \mu_2}\left(\mu_1 p_2 \sqrt{\cos\theta + \mu_2 \sin\theta} - \mu_2 p_1 \sqrt{\cos\theta + \mu_1 \sin\theta}\right)\right] \quad (E4)$$
$$+ b_{11} Tr \cos\theta,$$

and

$$u_2 = K_I \sqrt{\frac{2r}{\pi}} \text{Re}\left[\frac{1}{\mu_1 - \mu_2}\left(\mu_1 q_2 \sqrt{\cos\theta + \mu_2 \sin\theta} - \mu_2 q_1 \sqrt{\cos\theta + \mu_1 \sin\theta}\right)\right] \quad (E5)$$
$$+ b_{12} Tr \sin\theta,$$

for a loading specified by the combination $(K_I, T)$. In (E4) and (E5) $(p_k, q_k)$ are given by

$$p_k = b_{11}\mu_k^2 + b_{12} - b_{16}\mu_k, \quad (E6)$$

and

$$q_k = b_{12}\mu_k + \frac{b_{22}}{\mu_k} - b_{26}, \tag{E7}$$

where $k = (1,2)$.

### F.  Fracture mechanism maps and calculations for other metamaterial topologies

We have also reported fracture mechanism maps for three topologies in addition to the octet-truss. These calculations were carried out in a manner identical to the octet-truss. The asymptotic calculations of Section E require knowledge of the elastic constants and here we detail the elastic constants for the three additional metamaterial topologies.

#### F1.  The compound SC/BCC truss

The compound SC/BCC truss comprises two elementary trusses: a simple cubic (SC) truss that is penetrated by the inclined struts of the body-centered (BCC) truss as shown in Figs. 3f and 3g. Let $\ell$ denote the length of the strut of the SC truss (which is then also the size of the unit cell), $r_1$ the radius of the SC truss struts and $r_2$ the radius of the SC truss struts. Then the ratio of the volume fraction of the SC to BCC truss elements in the compound SC/BCC truss is

$$V_{\text{SC/BCC}} = \frac{\sqrt{3}}{4}\left(\frac{r_1}{r_2}\right)^2, \tag{F1}$$

while the relative density of the compound truss is

$$\bar{\rho} = \pi \left(\frac{r_2}{\ell}\right)^2 \left[3\left(\frac{r_1}{r_2}\right)^2 + 4\sqrt{3}\right]. \tag{F2}$$

Let $v_{SC} \equiv 100 V_{\text{SC/BCC}}/(1 + V_{\text{SC/BCC}})$ and $v_{BCC} \equiv 100 - v_{SC}$ denote the volume fraction in percent of the SC and BCC trusses in the compound truss. The notation used to describe the compound truss is then $v_{SC}\text{SC}/v_{BCC}\text{BCC}$ so that 60SC/40BCC denotes a compound truss with 60% by volume SC truss and 40% by volume BCC truss in the compound truss constructed by choosing $(r_1/r_2)^2 = 2\sqrt{3}$. In the co-ordinate system defined in Fig. 3f the elastic stress versus strain relation of this compound truss with cubic symmetry is given by

$$\begin{pmatrix}\varepsilon_{11}\\\varepsilon_{22}\\\varepsilon_{33}\\\gamma_{23}\\\gamma_{13}\\\gamma_{12}\end{pmatrix} = \frac{1}{\bar{\rho}E_s}\begin{pmatrix}13/3 & -2/3 & -2/3 & 0 & 0 & 0\\ & 13/3 & -2/3 & 0 & 0 & 0\\ & & 13/3 & 0 & 0 & 0\\ & & & 45/2 & 0 & 0\\ & \text{sym} & & & 45/2 & 0\\ & & & & & 45/2\end{pmatrix}\begin{pmatrix}\sigma_{11}\\\sigma_{22}\\\sigma_{33}\\\sigma_{23}\\\sigma_{13}\\\sigma_{12}\end{pmatrix}, \tag{F3}$$

where we have assumed the struts to be pin-jointed (as for the octet-truss discussed above this assumption is exceptionally accurate for stretching-governed trusses for $\bar{\rho} \leq 0.2$).

The compound truss formed by the choice $(r_1/r_2)^2 = 8/(3\sqrt{3})$ results in a 40SC/60BCC compound truss that is elastically isotropic with Young's modulus $E = \bar{\rho}E_s/6$ and Poisson's ratio $\nu = 1/4$.

*F2.   The chiral gyroid metamaterial*

This chiral metamaterial was analysed in *(18)* and here we report their results for the sake of completeness. For a gyroid comprising struts of length $b$ and radius $r_0$ the relative density is

$$\bar{\rho} = \frac{3\pi}{4\sqrt{2}} \left(\frac{r_0}{b}\right)^2, \tag{F4}$$

with the unit cell size related to strut length via $\ell = 2\sqrt{2}b$. The gyroid processes cubic symmetry and in the $e_i$ co-ordinate system of Fig. 3h the elastic stress versus strain relation can be written as

$$\begin{pmatrix} \varepsilon_{11} \\ \varepsilon_{22} \\ \varepsilon_{33} \\ \gamma_{23} \\ \gamma_{13} \\ \gamma_{12} \end{pmatrix} = \begin{pmatrix} 1/E & -\nu/E & -\nu/E & 0 & 0 & 0 \\ & 1/E & -\nu/E & 0 & 0 & 0 \\ & & 1/E & 0 & 0 & 0 \\ & & & 1/G & 0 & 0 \\ & \text{sym} & & & 1/G & 0 \\ & & & & & 1/G \end{pmatrix} \begin{pmatrix} \sigma_{11} \\ \sigma_{22} \\ \sigma_{33} \\ \sigma_{23} \\ \sigma_{13} \\ \sigma_{12} \end{pmatrix}, \tag{F5}$$

where

$$\frac{E}{E_s} = 0.426\bar{\rho}^2; \quad \frac{G}{E_s} = 0.329\bar{\rho}^2, \tag{F6}$$

and

$$\nu = \frac{1}{2} - \frac{E}{6K} \tag{F7}$$

with the bulk modulus $K$ given by

$$\frac{K}{E_s} = \frac{1}{9}\bar{\rho}. \tag{F8}$$

Thus, we observe that the chiral gyroid metamaterial is a bending-governed metamaterial under all loading conditions expect under pure hydrostatic loading with the bulk modulus being stretching-dominated.

*G.   Scaling laws and normalisation for the fracture mechanism map*

The finite element calculations and measurements show that the toughness for a given value of $\bar{T}$ scales with relative density $\bar{\rho}$ when the toughness is set by tensile strut fracture and with $\bar{\rho}^2$ when elastic strut buckling governs failure. Then, dimensional analysis dictates that for an octet-truss metamaterial made from an elastic-brittle parent material

$$K_{IC} = \begin{cases} \alpha(\bar{T})\bar{\rho}\sigma_f\sqrt{\ell} & \text{for } \bar{\rho} \geq \dfrac{\alpha}{\beta}\varepsilon_f \\[1em] \beta(\bar{T})E_s\bar{\rho}^2\sqrt{\ell} & \text{otherwise,} \end{cases} \tag{G1}$$

where $(\alpha, \beta)$ are functions of $\bar{T}$ that are numerically determined (e.g. $\alpha_0 \equiv \alpha(\bar{T} = 0) = 0.31$ and $\beta_0 \equiv \beta(\bar{T} = 0) = 0.33$ as detailed in the main text). Then recalling that $\sigma_f \equiv E_s\varepsilon_f$ it follows that

$$\frac{\overline{K}_{IC}}{\varepsilon_f} = \begin{cases} \alpha(\overline{T})\dfrac{\bar{\rho}}{\varepsilon_f} & \text{for } \bar{\rho} \geq \dfrac{\alpha}{\beta}\varepsilon_f \\[2ex] \beta(\overline{T})\left(\dfrac{\bar{\rho}}{\varepsilon_f}\right)^2 & \text{otherwise,} \end{cases} \quad (G2)$$

i.e. similar to the well established ideas for strength *(7)* the scaling laws for toughness too depend on $\bar{\rho}/\varepsilon_f$. Therefore $\overline{K}_{IC}/\varepsilon_f$ is a unique function of $(\bar{\rho}/\varepsilon_f, \overline{T})$ and this motivates the choice of axes and parameters used for the fracture mechanism maps in Fig. 3e, 3h and 3i.

**Supplementary Figures**

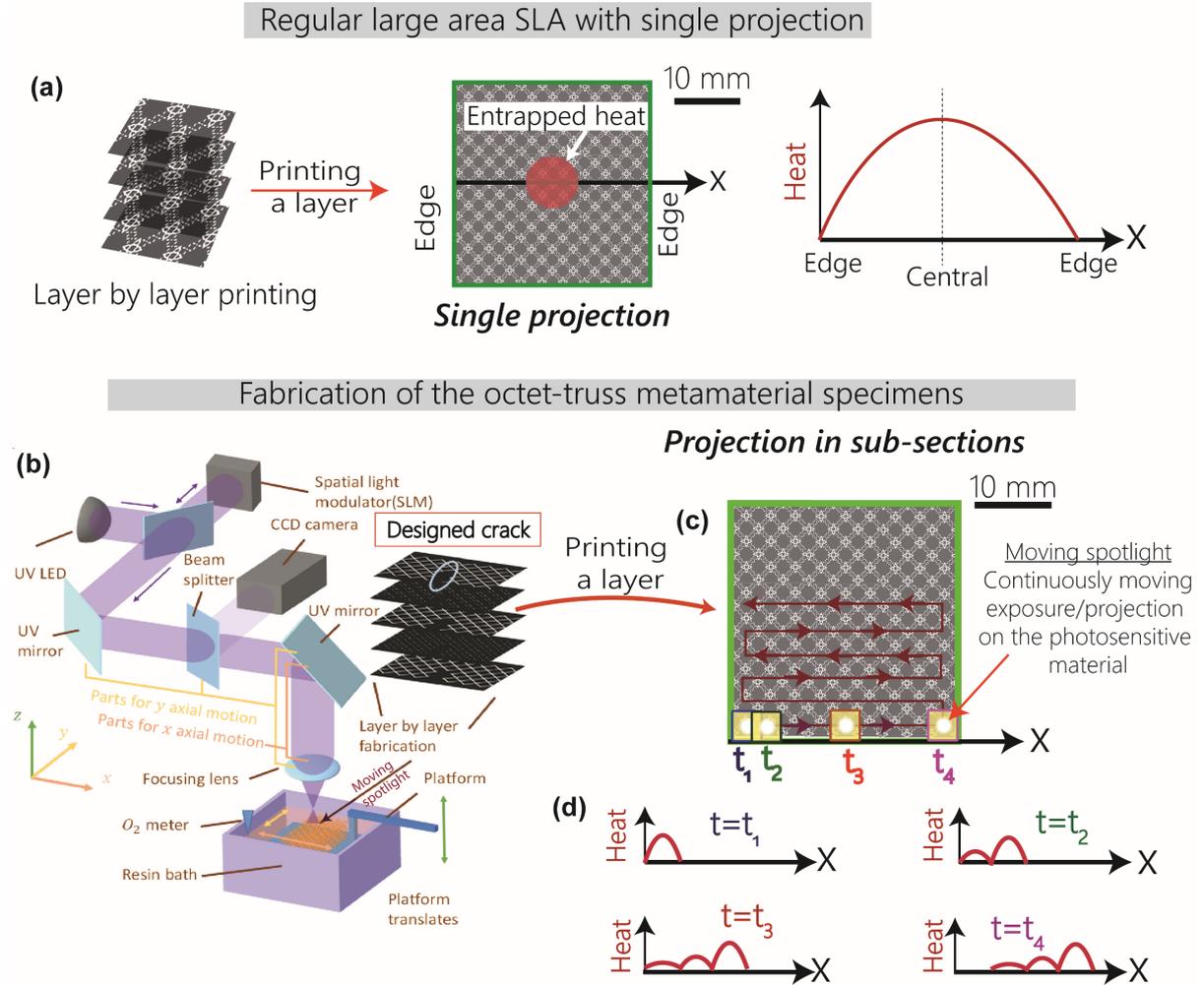

**Figure S1:** (a) Regular large area projection stereolithography (SLA) uses a single projection to print each layer resulting in entrapment of heat in the centre of the projection zone as depicted in the schematic. (b) The modified SLA system used to print the octet-truss metamaterial specimens incorporates an optical system comprising a moving spotlight so that (c) each layer in cured in sub-sections of the spotlight size with the entire layer cured by traversing the spotlight over the cross-section as shown schematically. This allows efficient heat dissipation. (d) Schematics depicting dissipation of heat over time ($t_1 < t_2 < t_3 < t_4$), where $t_i$ is the time that spotlight was focussed on location ($i$). Heat generated in the entire layer, particularly the central zone, is dissipated before the next layer is printed over it.

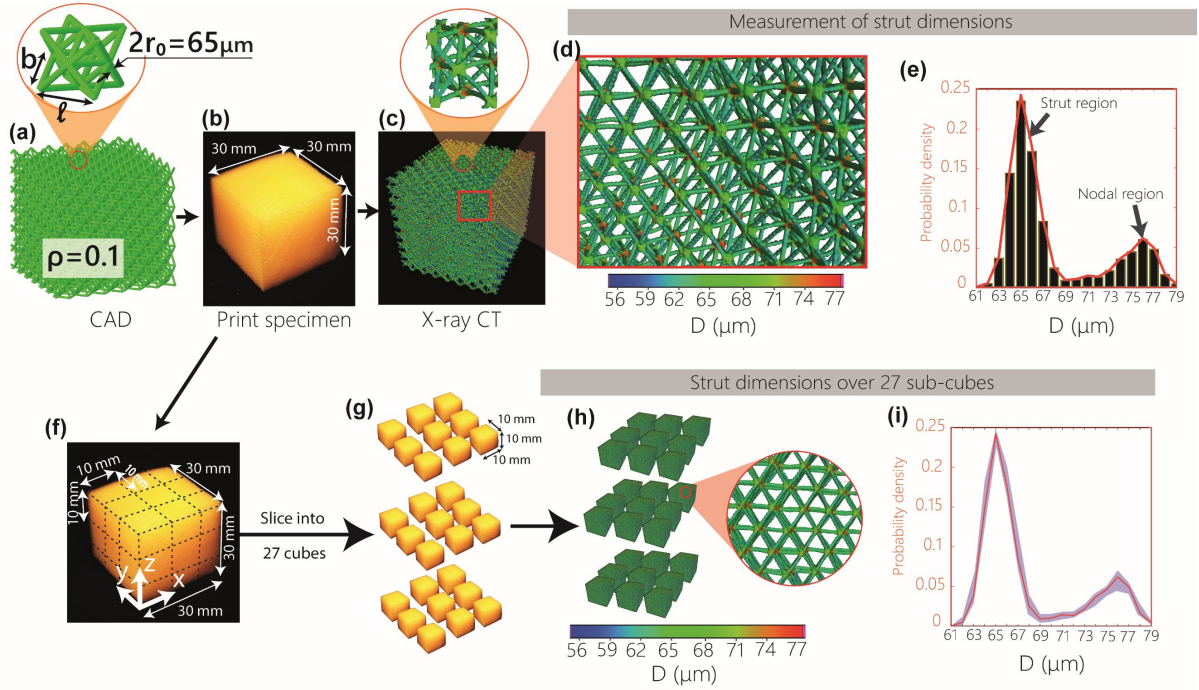

**Figure S2:** (a) CAD model for the $\bar{\rho} = 0.1$ specimen with strut diameter $2r_0 = 65$ μm along with (b) an optical image of the printed specimen. (c) XCT image of the entire specimen comprising ~64,000 unit cells with the inset showing a more detailed view of a single unit cell within the specimen. (d) A magnified view of a larger region in the specimen in (c) with each location within the solid material shaded by the diameter $D$ of the best-fit sphere at that location. (e) Histogram showing the distribution of diameters $D$ within the entire specimen with the left $y-$axis showing the count while the right $y-$axis giving the corresponding probability density. The two modes correspond to best-fit spheres within the struts and node regions as marked. (f) The printed specimen was divided into 27 sub-cubes as shown, (g) cut and (h) then each individual cube was separately imaged via XCT to determine the distribution of the diameter $D$ of the best-fit spheres as before. (i) The measured probability density function of $D$ over the 27 sub-cubes is shown by the shaded zone with the solid line giving the mean over the 27 separate sub-cubes.

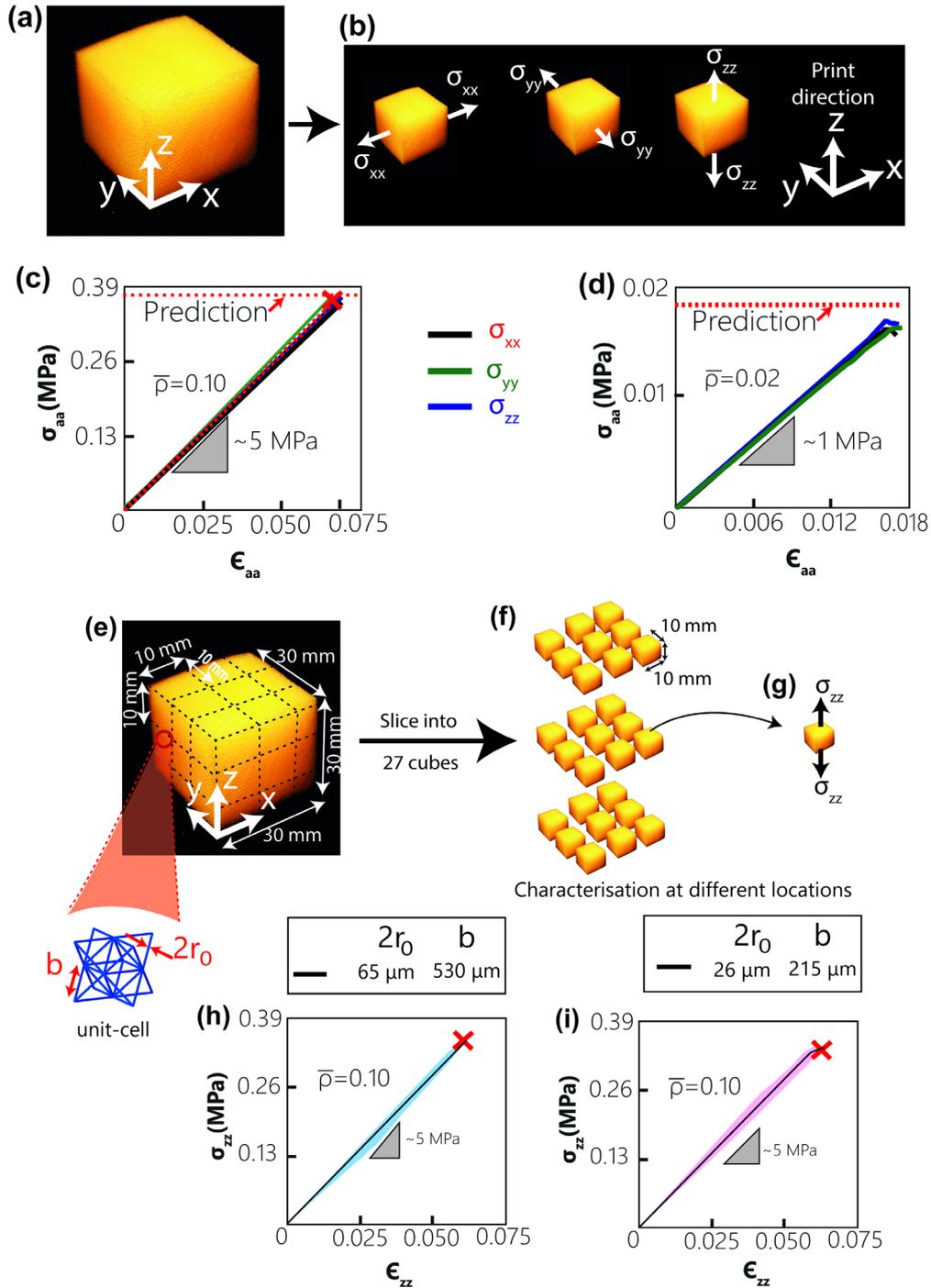

**Figure S3:** (a) Optical image of the printed specimen and (b) sketches depicting tensile testing along $x, y$ and $z-$ directions where $z$ is the print direction. The measured tensile responses for the (c) $\bar{\rho} = 0.1$ and (d) $\bar{\rho} = 0.02$ specimens in the $x, y$ and $z-$ directions. Predictions based on the measured properties of a single strut (Fig. S4) are included in (c) and (d). (e) The specimen cut into 27 sub-cubes and (f) uniaxial tensile tests were conducted on each cube along the (g) $z-$ direction. The measured responses are shown in (h) and (i) for the $\bar{\rho} = 0.1$ specimen with two different choices of unit cell dimensions. The shaded zone depicts the variation over the 27 sub-cubes with the solid line the mean measured response.

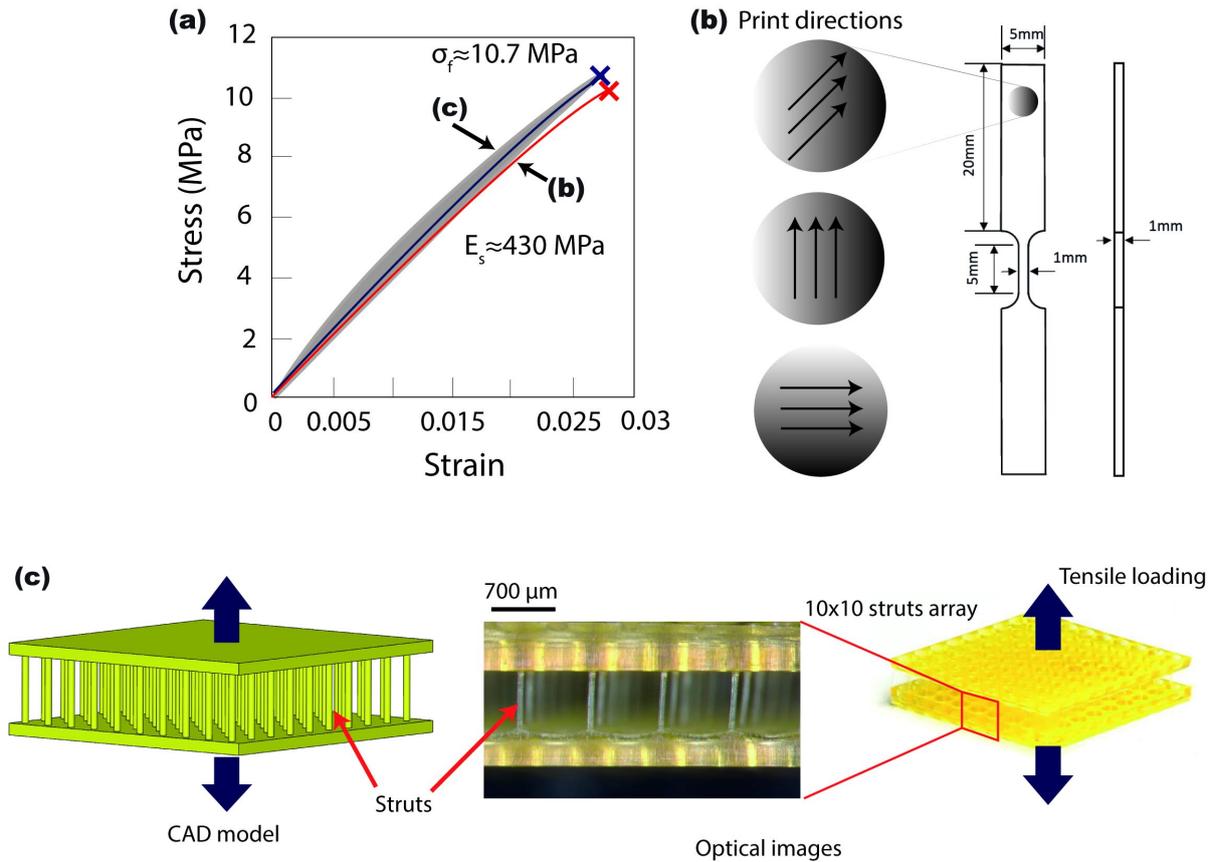

**Figure S4:** (a) A printed macro dogbone specimen (leading dimensions in the mm range) made from trimethylolpropane triacrylate (TMPTA) which is the parent material of the octet-truss metamaterials. The geometry, dimensions and print directions of three different specimens are shown. (b) The measured uniaxial tensile stress versus strain response of dogbone specimens printed in the three directions shown in inset of (a). The shaded thick grey line shows the variability over 3 separate tests for each print direction while the crosses mark the instant of brittle fracture of the specimens. Tensile response was also measured using (c) a printed $10 \times 10$ array of vertical struts of diameter $2r_0 \sim 45$ μm, height $b \sim 0.54$ mm and spacing between struts $S \sim 0.7$ mm as marked in the inset. We have included both (c) the CAD drawing used to print the array in order to clarify the geometry of the specimen, (d) an optical image of the full specimen used in these tests and (e) a more detailed optical view where we also indicate the print direction by arrows. In (b) we include the tensile material response extracted from both the dogbone and strut array measurements to show that the response is reasonably insensitive to the strut dimensions over the range of strut sizes used in the octet-truss specimens.

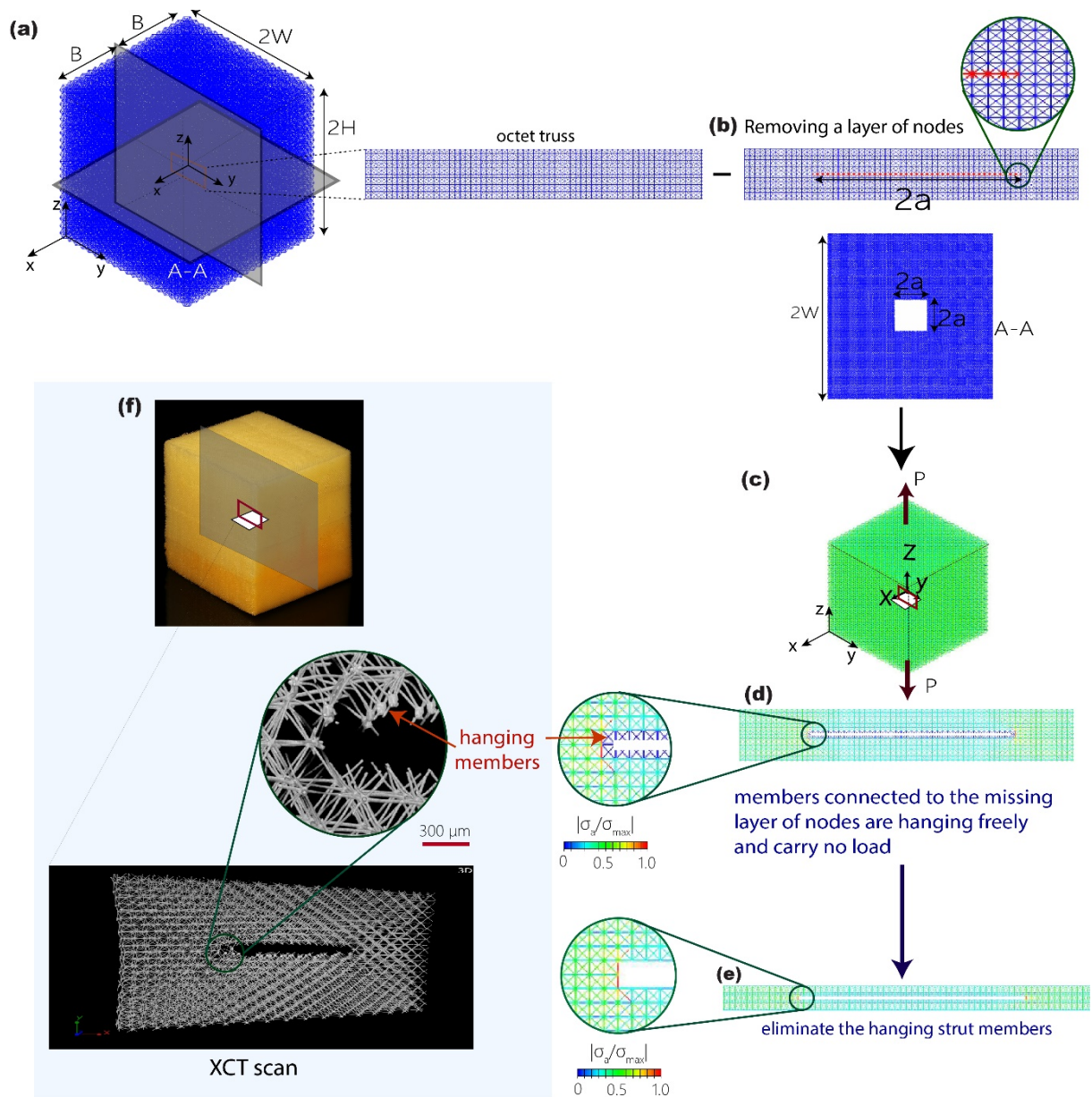

**Figure S5:** Sketches illustrating the creation of a "sharp" crack in the octet-truss metamaterial. We begin with (a) the uncracked octet-truss specimen and (b) remove a layer of nodes (shaded red) over a square region of size $2a \times 2a$ to create (c) the cracked specimen with (d) "hanging" struts over the crack flank. FE predictions of the distribution of axial stress within the struts around the crack are shown in (d) where the axial stresses $\sigma_a$ have been normalised by the maximum axial stress $\sigma_{max}$ within the entire specimen. The inset shows a zoomed in view over the crack flank to illustrate that the hanging struts over the crack flank carry no load. (e) FE predictions of the axial stress distributions with the hanging struts eliminated. This is equivalent to removal of a one layer of unit cells to form the crack. The stress distributions in (d) and (e) are identical confirming that the sharpest possible crack in the octet-truss corresponds to the removal of a one layer of unit cells. (f) Specimens manufactured with one layer of nodes removed and having hanging struts over the crack flank. Measurements confirmed that such specimens had an identical fracture response to specimens with a crack comprising one missing layer of unit cells.

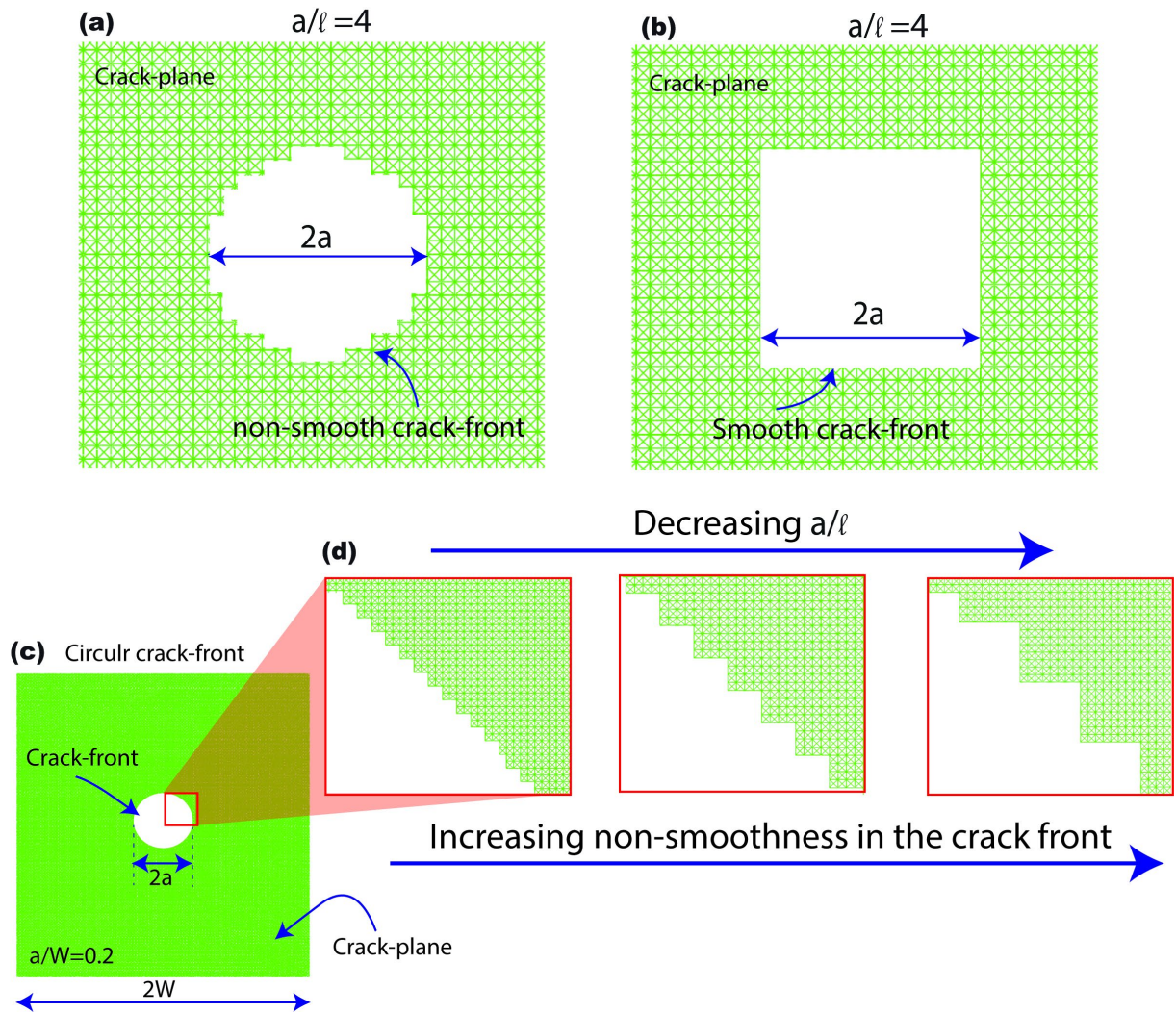

**Figure S6:** Sketches of (a) circular and (b) square crack geometries in the octet truss specimens for cracks of size $a/\ell = 4$ where $2a$ is the diameter of the circular crack or the side of the square crack and $\ell$ the cell size. The cubic geometry of the octet-truss unit cell implies that a "perfect" square crack geometry can be constructed. (c) The cubic unit cell implies that a circular crack shape can only be approximated via a jagged crack front. (d) The deviation from a perfect circular crack front becomes increasingly severe with decreasing $a/\ell$ shown for three values of $a/\ell$.

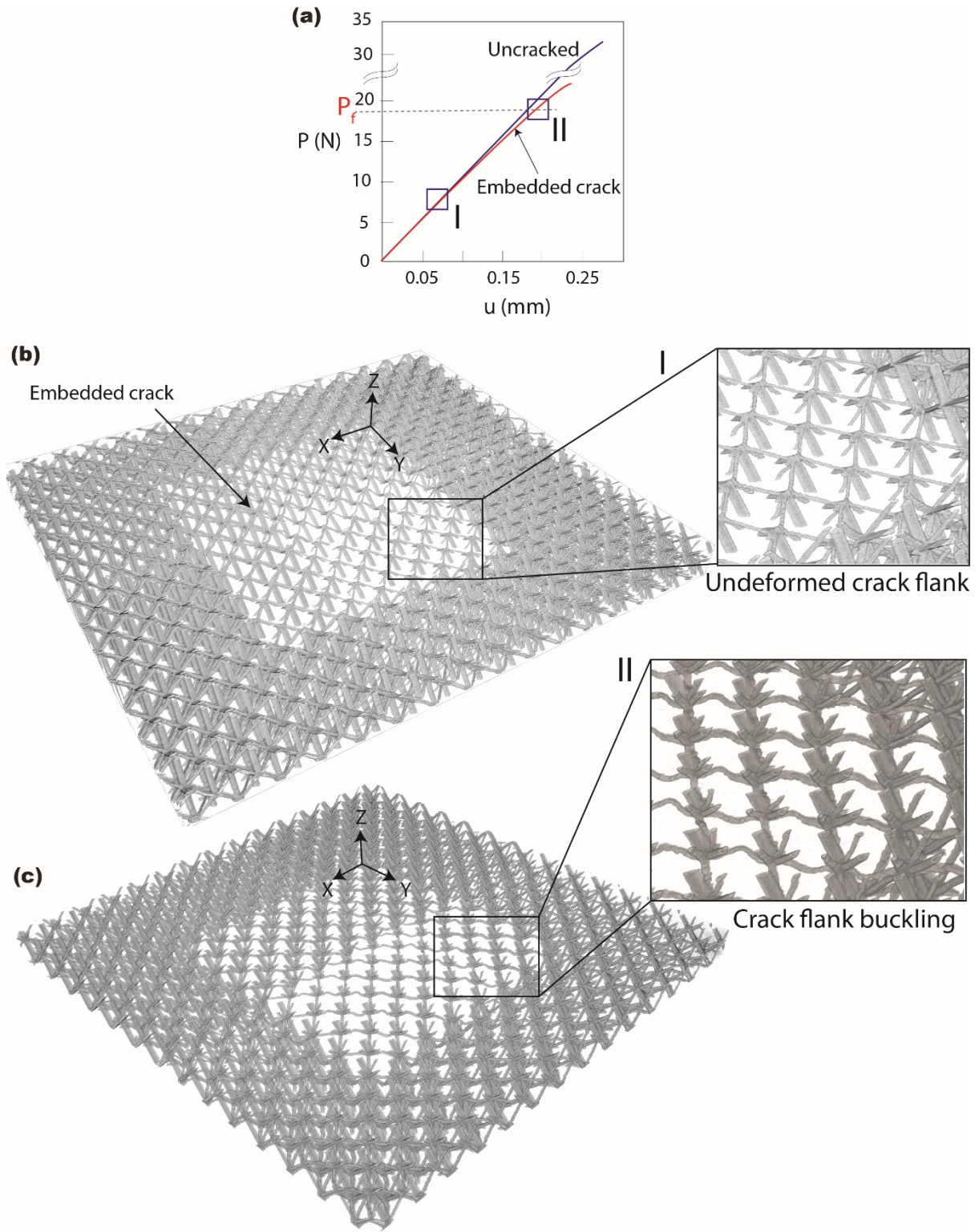

**Figure S7:** (a) Tensile load $P$ versus displacement $u$ response of the $\bar{\rho} = 0.03$ uncracked and cracked ($a/\ell = 4$) specimens where first failure occurs by elastic buckling of crack tip struts. The failure load $P_f$ is defined as the load at which the stiffness $dP/du$ reduces to 95% of its initial value. XCT images of the crack flank at (b) loading-stage I and (c) loading stage II marked in (a) to illustrate that $P_f$ corresponds to the initiation of crack flank buckling.

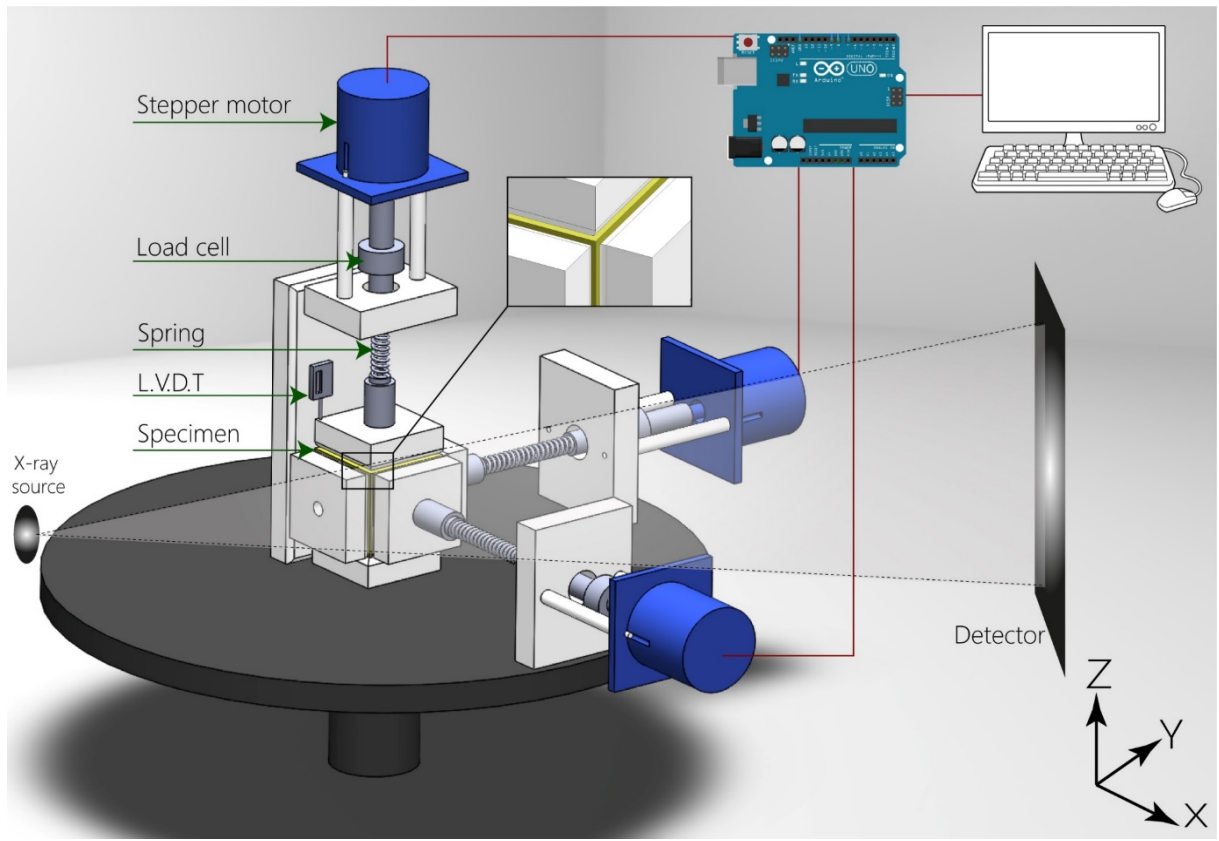

**Figure S8:** Sketch of the multi-axial loading setup and the in-situ X-ray imaging.

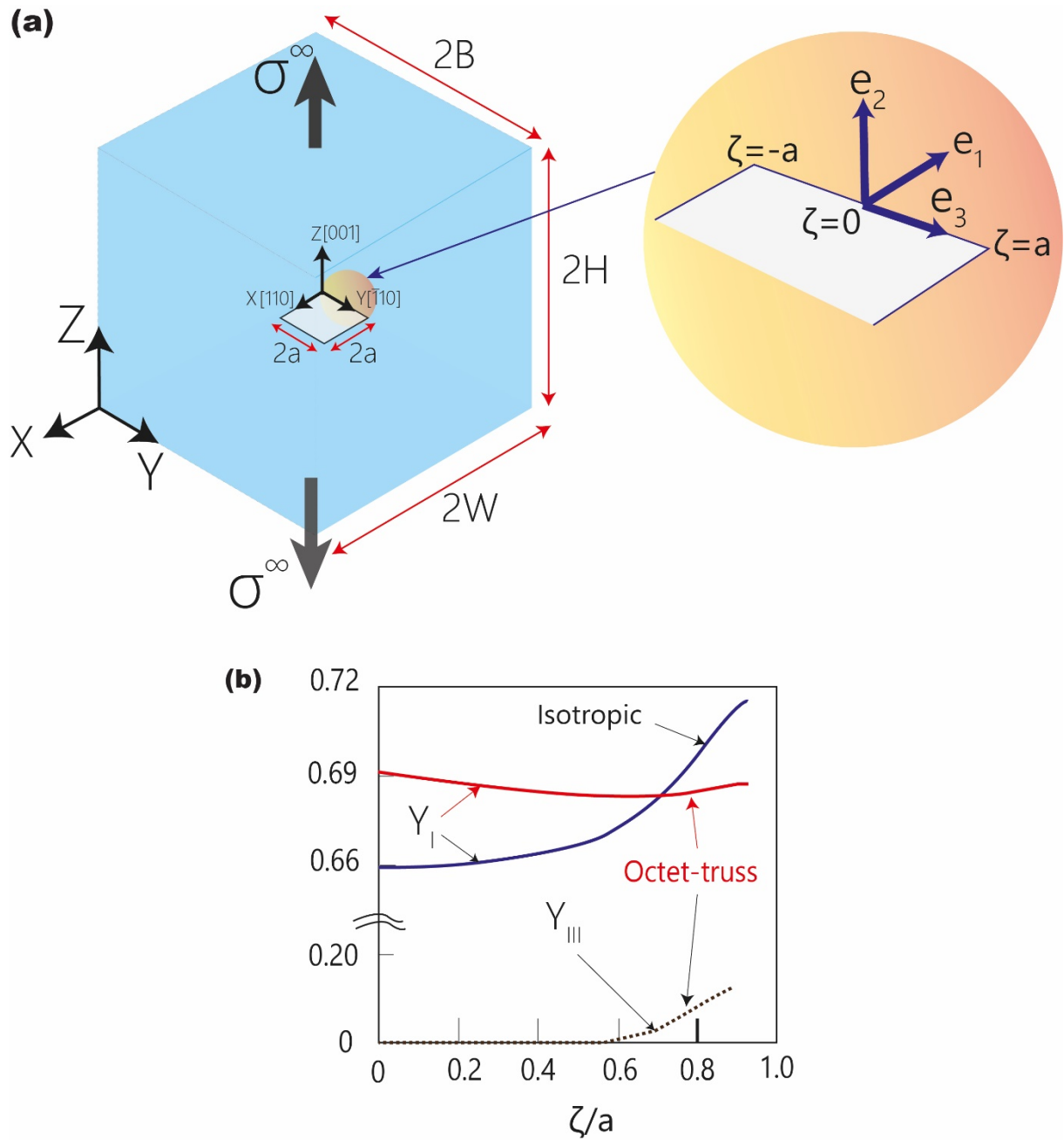

**Figure S9:** (a) Sketch illustrating orientation of the square crack in the 3D octet-truss specimen. The fcc crystallographic directions [110], [$\bar{1}$10] and [001] are also included to clarify the orientation. The inset shows the crack front co-ordinate system $e_i$ and the local co-ordinate $\zeta$ used to define a location along the crack front. (b) FE predictions of the calibration factors $Y_I$ and $Y_{III}$ as a function of the normalised position $\zeta/a$ along the crack front for crack growth in the plane of the crack and perpendicular to the crack front. These predictions use the cubic elastic constants of the octet-truss. Also included are predictions of $Y_I$ assuming an isotropic elastic medium with Poisson's ratio 0.3. All predictions are for a cracked geometry with $a/W = 0.2$.

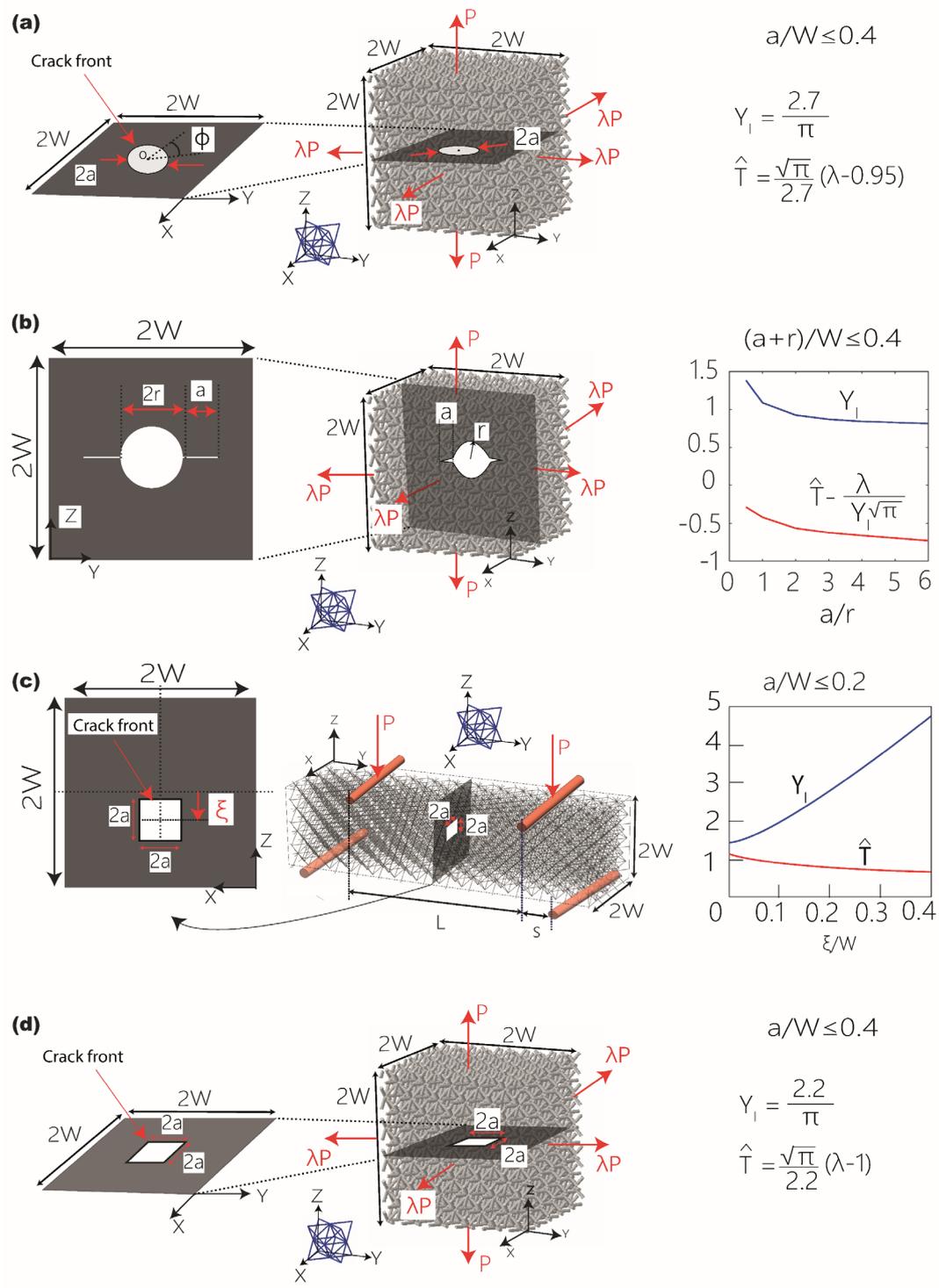

**Figure S10:** Calibration of $Y_I$ and $\hat{T}$ for the octet-truss metamaterial for a range of embedded crack geometries and loading configurations. (a) Penny-shaped crack of radius $a$ under multi-axial loading. (b) Circular crack emanating a distance $a$ from a spherical void of radius $r$ under multi-axial loading. (c) Square crack in a beam under 4-point bend situation and offset a distance $\xi$ from the neutral axis. (d) square crack of side $2a$ under multi-axial loading. In each case we have included the FE predictions of $Y_I$ and $\hat{T}$ along with the regime of validity of these predictions.

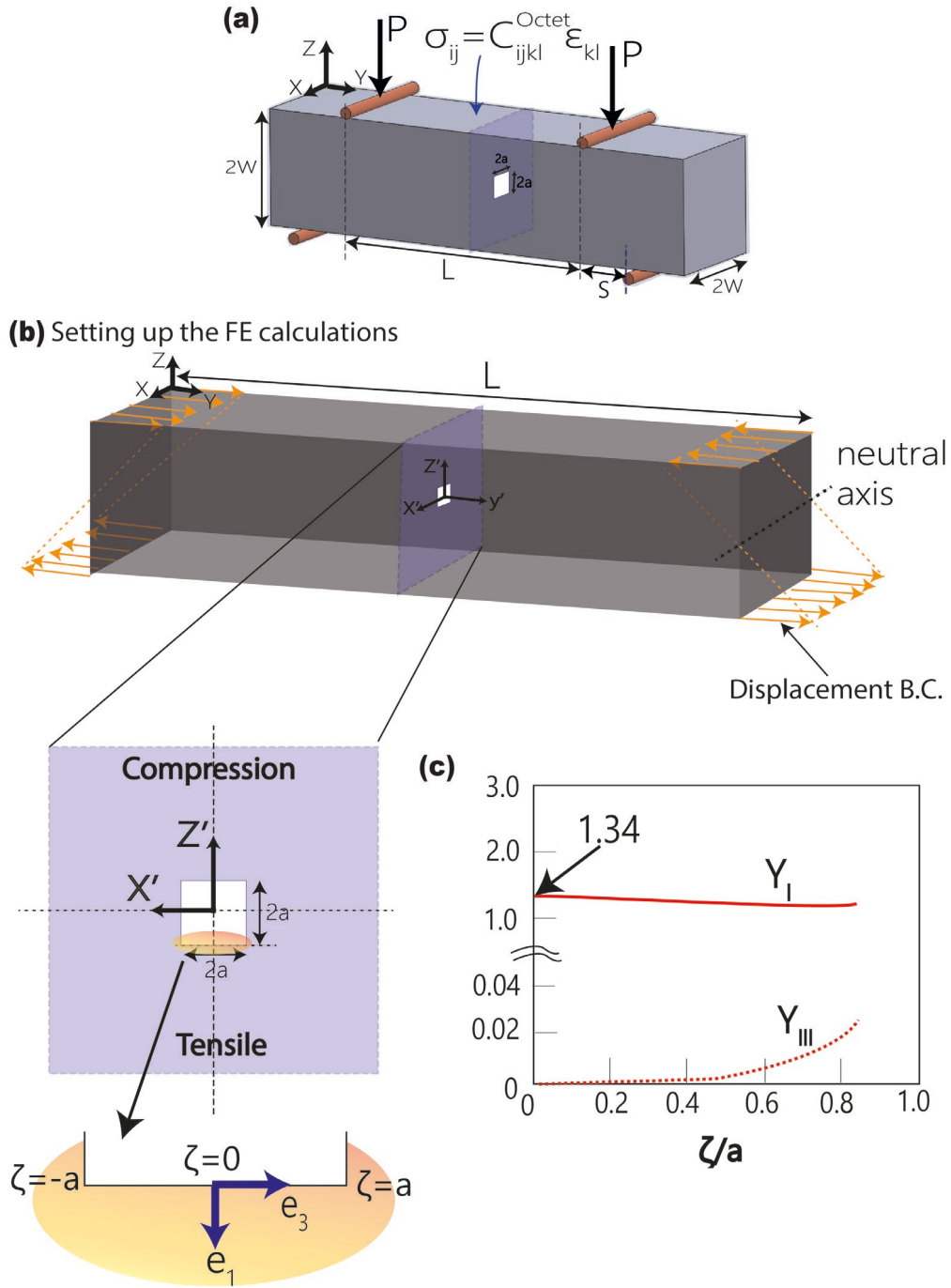

**Figure S11:** (a) Sketch of the cracked octet-truss metamaterial beam in 4-point bending modelled via a smeared-out continuum. (b) The FE calculations modelled a central section of length $L$ under pure moment loading which was applied via displacement boundary conditions on the beam ends as illustrated. The local co-ordinate system $(x', y', z')$ of this section is also marked. The inset shows the crack plane along with the local crack front co-ordinate system $e_i$ and the co-ordinate $\zeta$ used to define a location along the crack front. (c) FE predictions of the calibration factors $Y_I$ and $Y_{III}$ as a function of the normalised position $\zeta/a$ along the crack front for crack growth in the plane of the crack and perpendicular to the crack front. All predictions are for a cracked geometry with $a/W = 0.2$.

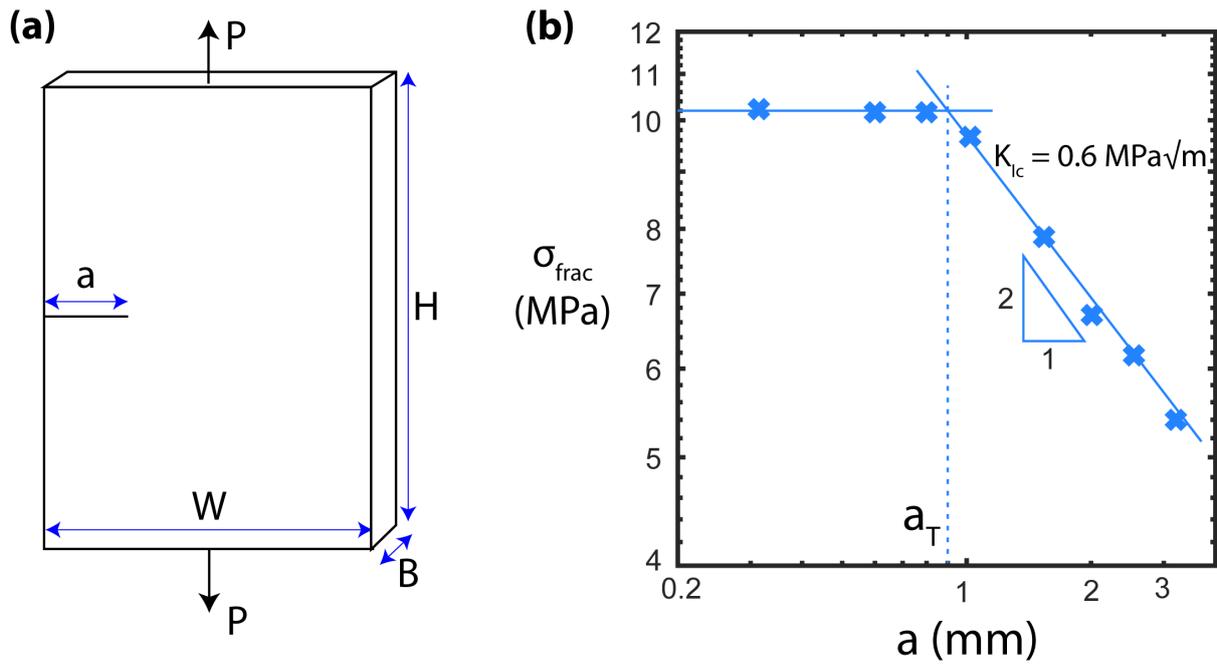

**Figure S12:** Sketch of the Single-Edge Notched Tension (SENT) specimen used for fracture toughness measurements of the parent solid TMPTA. (b) Measurements of the failure stress $\sigma_{\text{frac}} \equiv P_f/(WB)$ as a function of the crack size $a$ plotted on a log-log scale: $\sigma_{\text{frac}} \propto 1/\sqrt{a}$ for $a = a_T \geq 0.9$ mm with $\sigma_{\text{frac}} \approx \sigma_f \approx 11$ MPa for $a < a_T$ where $a_T \approx 0.9$ mm is the transition flaw size of the TMPTA.

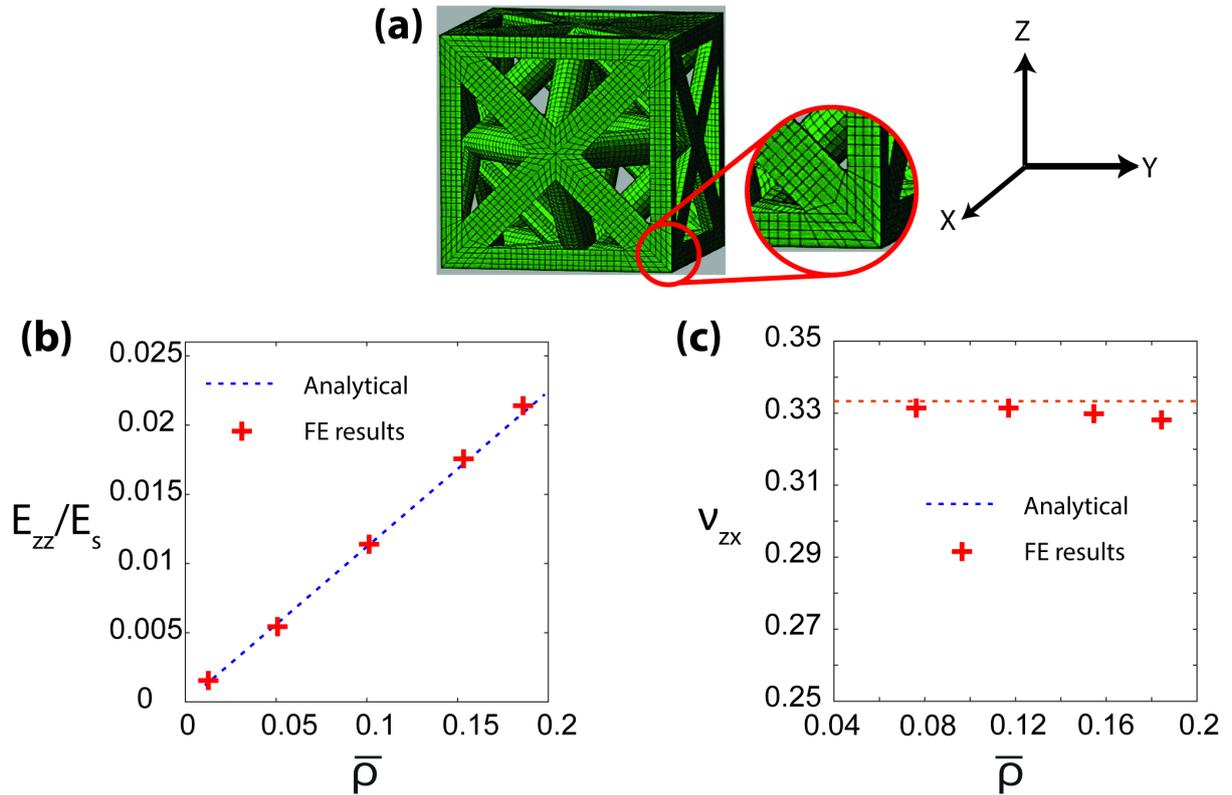

**Figure S13:** (a) The 3D FE model of an octet-truss unit cell constructed from the CAD model used in the SLA printing of the specimens. The FE discretion using solid 3D elements is shown along with the inset showing a detailed view of a node. Analytical and FE predictions of (b) Young's modulus $E_{zz}$ and (c) Poisson's ratio $\nu_{xz}$ as a function of relative density $\bar{\rho}$. Excellent agreement is observed between the FE and analytical predictions over the range of relative densities $\bar{\rho}$ considered in the present study.

**Table S1:** The maximum/minimum relative densities and corresponding crack sizes, strut lengths and diameters. The table shows the extremes of the ranges of the embedded crack samples investigated in this study.

| $\bar{\rho}$ | $a/\ell$ | Strut length $b(\mu m)$ | Strut diameter $2r_0(\mu m)$ |
|---|---|---|---|
| 0.02 | 4 | 530 | 29 |
|  | 10 | 212 | 12 |
| 0.10 | 4 | 530 | 65 |
|  | 10 | 215 | 26 |

**Table S2:** Parameters for the XCT scans.

| Scan protocol | Effective resolution (microns) | Beam Energy (W) | Binning | Number of 2D projections in 360° rotation | Exposure (fps) |
|---|---|---|---|---|---|
| Quick scan (for in-situ XCT) | ≤ 300 | 150-240 | 2 × 2 | <1000 | 4-12 |
| Long scan (for high quality imaging presented in this article) | 10 | 10 | 1 × 1 | ~3000 | 1 |

**Supplementary Movies**

Movies of 3D volumetric views of the specimens to illustrate the cracks and failure modes. The movies were constructed from the X-ray computed tomographic (XCT) imaging.

S1: The crack flank and front in the as-manufactured $\bar{\rho} = 0.03$ and $a/\ell = 4$ embedded crack specimen.

S2: Tensile fracture of the crack front struts in the $\bar{\rho} = 0.08$ and $a/\ell = 10$ embedded crack specimen loaded to $K_I = K_{IC}$.

S3: Buckling of struts over the crack flank in the $\bar{\rho} = 0.03$ and $a/\ell = 4$ embedded crack specimen loaded to $K_I = K_{IC}$.

S4: The $2B = 100$ unit cells, $\bar{\rho} = 0.10$ and $a/\ell = 10$ specimen with a through-thickness crack loaded to $K_I = K_{IC}$ showing the fracture of surface struts while struts within the specimen remained intact. In the movie, we show a slice consisting of the free surface and 5 layers in thickness instead of the entire 100 layers of thickness.